\documentclass[journal]{IEEEtran}
\usepackage[T1]{fontenc}
\usepackage{amsmath}
\usepackage[cmintegrals]{newtxmath}
\usepackage{bm}
\usepackage{cite}
\usepackage{booktabs}
\usepackage{graphicx}
\usepackage{multirow}
\usepackage{stfloats}
\usepackage{array} 

\usepackage[usenames,dvipsnames]{xcolor}

\definecolor{mygreen}{rgb}{0,0.5,0}
\definecolor{darkblue}{RGB}{0,0,150}
 \usepackage{draftwatermark}
 \SetWatermarkText{Preprint}

\ifCLASSINFOpdf
\else
\fi
\begin{document}
%
\title{A Speech Enhancement Algorithm based on Non-negative Hidden Markov Model and Kullback-Leibler  Divergence}
%
%
%

\author{Yang~Xiang,
        Liming~Shi,~\IEEEmembership{Student Member,~IEEE,}
        Jesper~Lisby~H\o jvang,
        Morten~H\o jfeldt~Rasmussen,
        and~Mads~Gr\ae sb\o ll~Christensen,~\IEEEmembership{Senior Member,~IEEE}
\thanks{Y. Xiang is an industrial PhD student. He is with the Audio Analysis Lab, CREATE, Aalborg University, Aalborg, Denmark and Capturi A/S, Aarhus, Denmark, e-mail: yaxi@create.aau.dk}
\thanks{L. Shi and M. G. Christensen are with the Audio Analysis Lab, CREATE, Aalborg University, Aalborg, Denmark, e-mail:\{ls,mgc\}@create.aau.dk}
\thanks{J. L. Højvang and M. H. Rasmussen are with Capturi A/S, Aarhus, Denmark, e-mail: \{jlh,mhr\}@capturi.com} 
\thanks{This  work  was  supported  by  Capturi  and  Innovation  FundDenmark (Grant No.9065-00046).}
 }

\raggedbottom

%
%

\markboth{Journal of \LaTeX\ Class Files,~Vol.~14, No.~8, June~2020}%
{Shell \MakeLowercase{\textit{et al.}}: Bare Demo of IEEEtran.cls for IEEE Journals}
%



\maketitle

\begin{abstract}
In this paper, we propose a novel supervised single-channel speech enhancement method combing the the Kullback-Leibler divergence-based non-negative matrix factorization (NMF) and hidden Markov model (NMF-HMM). With the application of HMM, the temporal dynamics information of speech signals can be taken into account. In the training stage, the sum of Poisson, leading to the KL divergence measure, is used as the observation model for each state of HMM. This ensures that a computationally efficient multiplicative update can be used for the parameter update of the proposed model. In the online enhancement stage, we propose a novel minimum mean-square error (MMSE) estimator for the proposed NMF-HMM. This estimator can be implemented using parallel computing, saving the time complexity. The performance of the proposed algorithm is verified by objective measures. The experimental results show that the proposed strategy achieves better speech enhancement performance than state-of-the-art speech enhancement methods. More specifically, compared with the traditional NMF-based speech enhancement methods, our proposed algorithm achieves a 5\% improvement for short-time objective intelligibility (STOI) and 0.18 improvement for perceptual  evaluation  of  speech  quality (PESQ).
\end{abstract}


\begin{IEEEkeywords}
speech enhancement, non-negative matrix factorization (NMF), hidden Markov model (HMM), minimum mean-square error (MMSE), Kullback-Leibler (KL) divergence.
\end{IEEEkeywords}

%
\IEEEpeerreviewmaketitle

\section{Introduction}
%
%
%
%
\IEEEPARstart{S}{ingle-channel} speech enhancement technology has been widely used in our daily lives, such as speech coding, teleconferencing, hearing aids, mobile
communication, and automated robust speech recognition (ASR) \cite{li2014overview,loizou2013speech}. In general, the purpose of speech enhancement is to remove background noise from noisy speech while preserving clean speech. It aims to improve the quality and intelligibility of noisy speech \cite{xu2013experimental}. Currently, single-channel speech enhancement is an active topic of research.

During the past decades, many different monaural speech enhancement approaches have been proposed \cite{loizou2013speech,cohen2008spectral}. In an environment with additive noise, the simplest approach to conduct speech enhancement is the spectral subtraction algorithm \cite{boll1979suppression}, which subtracts the estimated noise spectrum from the observed noisy speech in order to acquire the desired clean speech. Additionally, other unsupervised methods such as signal subspace algorithm \cite{christensen2016experimental,jensen2015noise,ephraim1995signal,jabloun2003incorporating}, Wiener
filtering \cite{lim1978all}, minimum mean-square error (MMSE) spectral amplitude estimator \cite{ephraim1984speech}, and log-MMSE spectral amplitude estimator \cite{ephraim1985speech} are effective strategies to conduct speech enhancement when the noise is stationary. These methods have low computational complexity, so they have been widely applied in various areas. However, these approaches cannot always achieve satisfactory performance for non-stationary noise and usually introduce musical noise because they do not make best use of the prior information of speech and noise \cite{hussain2007nonlinear}. Moreover, most of unsupervised methods are based the statistical properties of the speech and noise signals. However, it is difficult to meet these properties in real-world noisy scenarios \cite{xu2014regression}.

Therefore, supervised speech enhancement approaches have been developed. For instance, Srinivasan \cite{srinivasan2007codebook} proposed a codebook-driven speech enhancement algorithm for non-stationary noise. In this work, the auto-regressive (AR) spectrum shape codebooks of speech and noise were pre-trained. In the enhancement stage, the codebooks could be used to build a Wiener filter to conduct speech enhancement. Inspired by this research, many other codebook-based speech enhancement approaches have been developed \cite{kavalekalam2018online,he2016multiplicative}. Furthermore, auto-regressive hidden Markov model (ARHMM) \cite{zhao2007hmm,deng2015sparse} is also an effective supervised speech enhancement method because it considers the temporal information of speech signal.

In recent years, with the advance of hardware and deep learning technologies \cite{bengio2009learning,hinton2006fast}, deep neural networks (DNNs) have significantly promoted to the development of speech enhancement \cite{wang2018supervised}. These methods usually rely on fewer assumptions \cite{xu2013experimental,xu2014regression,wang2018supervised} between noise and clean speech, so they have huge potential to achieve better speech enhancement performance. Xu \cite{xu2013experimental,xu2014regression} applied a  feedforward multilayer perceptron (MLP) to map log-power spectrum (LPS) features of clean speech given noisy LPS input. The enhanced speech could be obtained directly by waveform reconstruction. Compared with the MMSE estimator \cite{ephraim1985speech}, this method achieved better performance in various noisy environments. Wang \cite{wang2014training,narayanan2013ideal} also utilized the MLP to estimate the ideal ratio mask (IRM) and ideal binary mask (IBM) to conduct speech enhancement, which also achieved satisfactory performance. Motivated by this work, different DNN structures have been used to conduct speech enhancement, such as fully convolutional neural network (FCN) \cite{park2016fully}, deep recurrent neural networks \cite{jacobsson2005rule,huang2015joint} and generative adversarial networks (GANs)\cite{goodfellow2014generative,pascual2017segan}. These methods could help  ASR systems achieve higher recognition accuracy in noisy environments. However, generalization is always a problem that needs to be considered for these DNN-based algorithms \cite{kolbaek2016speech,xiang2020parallel}.

Non-negative matrix factorization (NMF)-based \cite{lee1999learning,lee2001algorithms,shimada2019unsupervised} speech enhancement algorithms can be also viewed as a kind of supervised speech enhancement method. In \cite{grais2011single}, a mask-based NMF speech enhancement method was proposed. In the offline stage, the basis matrix of clean speech and noise was trained. In the enhancement stage, the activation matrix could be acquired by combining the trained basis matrix and noisy signal. Then, the mask was estimated to conduct speech enhancement. Additionally, an NMF-based denoising scheme was described in \cite{wilson2008regularized,nie2016exploiting}, which added a heuristic term to the cost function, so the NMF coefficient could be adjusted according to the long-term levels of signals. A parametric NMF method for speech enhancement was proposed in \cite{kavalekalam2018online}. This method applied the AR coefficient and codebook to build the basis matrix. This strategy effectively improved speech intelligibility. Moreover, some DNN-based NMF methods represent an effective strategy to conduct speech enhancement \cite{bando2018statistical}. In general, the basis matrix could be acquired using the traditional NMF method and the activation matrix could be estimated by applying the DNN \cite{kang2014nmf}, which improved the accuracy of the estimated activation matrix. Thus, it could achieve higher perceptual evaluation of speech quality (PESQ) \cite{rix2001perceptual} and short-time objective intelligibility (STOI) \cite{taal2011algorithm} scores than traditional NMF-based speech enhancement methods. The combination of DNN and NMF could also help the ASR system achieve a lower word error rate (WER) in noisy environments. In \cite{vu2016combining}, a DNN-NMF-based method achieved excellent performance in the CHiME-3 challenge. To capture temporal information, some HMM-based NMF speech enhancement methods have been proposed. Mohammadiha \cite{mohammadiha2013supervised} proposed a supervised and unsupervised NMF speech enhancement method. In \cite{mohammadiha2013supervised}, HMM is used for modeling the temporal change of different noise types. In \cite{mysore2010non}, a non-negative factorial HMM was used to model sound mixtures and showed superior performance in source separation tasks. In \cite{wang2016dnn}, an HMM-DNN NMF speech enhancement algorithm was proposed, which applied clustering method to acquire the HMM-based basis matrix and used Viterbi algorithm to obtain the ideal state label for the DNN training. In the enhancement stage, the DNN was used to find the corresponding state to conduct speech enhancement. This strategy achieved satisfactory speech enhancement performance.

Inspired by previous studies, in this paper, we propose a novel NMF-HMM speech  enhancement method based on the Kullback-Leibler (KL) divergence. This method uses the HMM to capture the temporal dynamics of speech signal. Moreover, we use the sum of Poisson distribution as the state-conditioned likelihood for the HMM, rather than the general gaussian mixture Model (GMM), because the sum of Poisson distribution leads to the KL divergence measure, which is a mainstream measure in NMF, and its parameter update rule is identical to the multiplicative update rule. This ensures that the parameter update is computationally efficient during the training stage. In the enhancement stage, we propose a novel NMF-HMM-based MMSE estimator to conduct the online speech enhancement. Another benefit of the proposed algorithm is that the  activation matrix can be updated by parallel computing in the online stage. This can effectively reduce computational time. The proposed method is evaluated by PESQ and STOI.

The rest of this paper is organized as follows. First, we will briefly review the general NMF-based speech enhancement method with KL divergence in Section II. The HMM-based signal model will be introduced in Section III and the offline parameter learning will be explained in Section IV. The details of proposed MMSE estimator and online speech enhancement process will be given in Section V. The experimental comparison and analysis of results will be illustrated in Section VI and we will draw conclusions in Section VII.

\section{NMF-based speech enhancement method with KL divergence}
In this section, we will briefly review the NMF-based speech enhancement with KL divergence. Assuming additive noise, the noisy signal model can be expressed as
\begin{equation}
  y(t) = s(t)+m(t),
  \label{eq1}
\end{equation}
 where \(y(t)\), \(s(t)\) and \(m(t)\) denote the noisy signal, clean speech and noise, respectively, and $t$ is the time index. Using \eqref{eq1}, the short-time Fourier transform (STFT) of \(y(t)\) can be written as
   \begin{equation}
  Y(f,n) = S(f,n)+M(f,n),
  \label{eq2}
\end{equation}
  where $Y(f,n)$, \(S(f,n)\), and \(M(f,n)\) denotes the frequency spectrums of \(y(t)\), \(s(t)\), and \(m(t)\), respectively. Here, $f \in [1, F]$ and $n \in [1, N]$ denote the frequency bin and time frame indices, respectively. Collecting $F$ frequency bins and $N$ time frames, {we define the magnitude spectrum matrices $\mathbf{Y}_N$, $\mathbf{S}_N$ and $\mathbf{M}_N$, where  ${\mathbf{Y}}_N=[\mathbf{y}_1,\cdots, \mathbf{y}_n, \cdots, \mathbf{y}_N]$ and $\mathbf{y}_n=[|Y(1,n)|, \cdots, |Y(f,n)|,\cdots, |Y(F,n)|]^T$, $\mathbf{s}_n$ and $\mathbf{m}_n$ are defined similarly to $\mathbf{y}_n$. And $\mathbf{S}_N$ and $\mathbf{M}_N$ are defined similarly to $\mathbf{Y}_N$. Additionally, we assume that there is $\mathbf{Y}_N=\mathbf{S}_N+\mathbf{M}_N$}. The classical NMF-based speech enhancement has two stages: training and enhancement. In the training stage, the clean speech basis matrix \(\overline{\bf{W}}\) and noise basis matrix \(\ddot{\bf{W}}\) are trained using clean speech and noise databases, respectively. Many cost functions have been proposed for NMF, such as the KL divergence \cite{lee2001algorithms}, IS divergence \cite{fevotte2009nonnegative}, $\beta$ divergence and Euclidian distance \cite{fevotte2011algorithms}. In this paper, we focus on using the KL divergence measure. {There are two reasons for this choice. First, compared with other types of cost functions, the best speech enhancement  performance can be achieved using the KL divergence-based NMF with the magnitude spectrum \cite{fitzgerald2009use}. Second, the efficient multiplicative update (MU) rule of the KL divergence-based NMF can be also derived statistically using the  expectation maximization (EM) algorithm \cite{cemgil2009bayesian}.} For two matrices $\bf{B}$ and $\hat{\bf{B}}$, the KL divergence measure is defined as
     \begin{equation}
 {{\rm KL}(\bf{B|\hat{B})}} = \sum_{i,j} (b_{i,j} {\rm log}(b_{i,j}/ \hat{b}_{i,j})-b_{i,j}+\hat{b}_{i,j}),
  \label{eq3}
\end{equation}
where $b_{i,j}$ and ${\hat b}_{i,j}$ denote the elements from the $i^{\mathrm{th}}$ row and $j^{\mathrm{th}}$ column of the matrices $\bf{B}$ and $\bf\hat{B}$, respectively. Using speech basis matrix training as an example, the {the cost function of the KL divergence-based} NMF for training $\overline{\bf{W}}$ can be written as
{\begin{equation}
  (\overline{\bf{W}},\overline{\bf{H}}) = {\mathop{\arg\min}_{\overline{\bf{W}},\overline{\bf{H}}}}{\ } {{\mathrm {KL}}(\mathbf{S}_N|\overline{\bf{W}}\times \overline{\bf{H}})}.
  \label{eq4}
\end{equation}}
The noise basis matrix training is similar to the speech basis matrix training.   In \cite{lee2001algorithms}, it is derived that $\overline{\bf{W}}$ and $\overline{\bf{H}}$ can be obtained {iteratively} using the following multiplicative update rules:
{ \begin{equation}
   {\overline{\bf{W}}} \leftarrow  {\overline{\bf{W}}} \odot {\cfrac{\cfrac{\mathbf{S}_N}{\overline{\bf{W}}\times\overline{\bf{H}}}\overline{\bf{H}}^\mathit{T}}{\bf{1}{\overline{\bf{H}}}^\mathit{T}}}
  \label{eq5},
\end{equation}
     \begin{equation}
   {\overline{\bf{H}}} \leftarrow  {\overline{\bf{H}}} \odot {\cfrac{\overline{\bf{W}}^\mathit{T}\cfrac{\mathbf{S}_N}{\overline{\bf{W}}\times\overline{\bf{H}}}}{\overline{\bf{W}}^\mathit{T}\bf{1}}},
  \label{eq6}
\end{equation}
}
where \(\odot\) and all divisions are element-wise multiplication and division operations, respectively,  $\bf{1}$ is a matrix of ones with the same size as $\mathbf{S}_N$. In the enhancement stage, the noisy speech basis matrix $\bf{W}$ can be constructed by concatenating the speech and noise basis matrices, i.e.,  ${{\bf{W}}=[\overline{\bf{W}},\ddot{\bf{W}} ]}$. The activation matrix ${\bf{H}}$ of the noisy speech can be estimated iteratively using \eqref{eq6} but replacing $\mathbf{S}_N$, $\overline{\bf{W}}$ and $\overline{\bf{H}}$ in \eqref{eq6} with $\mathbf{Y}_N$, $\mathbf{W}$ and $\mathbf{H}$, respectively.  The enhanced signal can be obtained using various algorithms \cite{grais2011single,wilson2008regularized, mohammadiha2013supervised,mysore2010non}.  One popular approach is to use the following Wiener-filtering like spectral gain $\mathbf{g}_n^{\text{NMF}}$ function:
\begin{align}
\mathbf{g}_n^{\text{NMF}}&= \frac{\overline{\mathbf{W}}\,\overline{\mathbf{h}}_{n}}{{\overline{\mathbf{W}}\,\overline{\mathbf{h}}_{n}}+{\ddot{\mathbf{W}}\,\ddot{\mathbf{h}}_{n}}}, \label{nmfgain}\\
\mathbf{h}_n&=\left[\overline{\mathbf{h}}_n^{T}, \ddot{\mathbf{h}}_n^{T}\right]^{T}  \nonumber \\
&=\arg \min_{\mathbf{h}_n } \mathrm{KL}(\mathbf{y}_n|\mathbf{W}\mathbf{h}_n), \label{onlineh}
\end{align}
where \eqref{onlineh} can be solved iteratively by using \eqref{eq6}. Apart from the gradient descent derivation of the MU update rules \eqref{eq5} and \eqref{eq6} presented in \cite{lee2001algorithms}, it is further shown in \cite{cemgil2009bayesian} that the MU update rules can be derived from a statistical perspective. More specifically, the KL divergence-based NMF can be motivated from the following hierarchical statistical model:
     \begin{equation}
  { {\bf{S}}_N = \sum_{k=1}^K {\bf{C}}(k),}
  \label{eq7}
\end{equation}
     \begin{equation}
   {c_{f,n}(k)} \sim \mathcal{PO} (c_{f,n}(k);\overline{W}_{f,k}\overline{H}_{k,n}),
  \label{eq8}
\end{equation}
where $\small{\mathcal{PO}(x;\lambda)=\cfrac{\lambda^{x}e^{-\lambda}}{\Gamma(x+1)}}$ is the Poisson distribution,  $\Gamma(x+1)=x! $ denotes the gamma function for positive integer $x$, $K$ denotes the number of basis vectors, $\mathbf{C}(k)$ is the latent matrix and $c_{f,n}(k)$ denotes the element of $\mathbf{C}(k)$ in the $f^\mathrm{th}$ row and $n^\mathrm{th}$ column.  Note that $ {c_{f,n}(k)}$ is assumed to have a Poisson distribution which can only be used for discrete variable. However, in practice, this hierarchical statistical model is not limited for discrete variables since the gamma function for continuous variable can be used to replace the factorial calculation \cite{cemgil2009bayesian}.  It has been shown in \cite{cemgil2009bayesian} that the iterative update of the parameters $\overline{\bf{H}}$ and $\overline{\bf{W}}$ using the expectation maximization (EM) algorithm is identical to the multiplicative update rules shown in (\ref{eq5}) and (\ref{eq6}).

{One of the advantages of the classical NMF-based method for speech enhancement is that the computational efficient MU rules can be applied. However, the temporal dynamical aspects of speech and noise are not taken into account. To incorporate the temporal dynamical information of audio signal, the HMM model is used in \cite{mysore2010non} for source separation. However, the parameter update rules are computational complex. Moreover, only off-line enhancement approach is presented.  In speech enhancement application, to consider the change of the noise types over time, the HMM model is used in \cite{mohammadiha2013supervised} to model the transition of the noise types over time. In this paper, we propose a NMF-based speech  enhancement algorithm using the HMM to take the temporal aspects of both the speech and noise into account. Moreover, an online MMSE estimator for speech enhancement is derived.}

\section{HMM-based signal models with the KL divergence}

{In this section, we present the proposed signal models, i.e., the speech and noise signal models, and the noisy signal model.

\subsection{Speech and Noise Signal Models}

In this work, the same signal model is used for both the clean speech and the noise signal, so we will illustrate them using only the clean speech signal. Additionally, we use the overbar ($\overline{\cdot}$) and double dots ($\ddot{\cdot}$) to represent the clean speech and the noise, respectively.  To consider the temporal dynamics information of the speech and noise, we use the HMM. Following the conditional independence property of the standard HMM \cite{baum1972inequality}, the likelihood function can be expressed as follows:
     \begin{equation}
   p({\bf{S}}_N;{\boldsymbol \Phi}) =  \sum_{\bf{\overline{x}}_N}\prod_{n=1}^N p({\bf s}_n|\overline {x}_n)p(\overline {x}_n|\overline {x}_{n-1}),
  \label{eq9}
\end{equation}
where ${\bf{\overline{x}}_N}=[\overline {x}_1,\cdots, \overline {x}_n,\cdots,\overline {x}_N]^T$ is a collection of states, $\overline {x}_n \in \{1,2,\cdots,\overline J\}$ denote the state at the $n^\mathrm{th}$ frame and $\overline{J}$ denotes the total number of states.  $p(\overline {x}_n|\overline {x}_{n-1})$ denotes the state transition probability from state $\overline {x}_{n-1}$ to $\overline {x}_{n}$ with $p(\overline {x}_1|\overline {x}_{0})$ being the initial state probability. $p({\bf S}_n|\overline {x}_n)$ is the state-conditioned likelihood function, $\boldsymbol \Phi$ is a collection of modeling parameters. Next, we describe the state transition probability and the state-conditioned likelihood function, respectively, for the proposed signal model.
\newline
\textbf{The state transition probability}  $p(\overline {x}_n|\overline {x}_{n-1})$: Following the standard HMM, we use the first-order-Markov-chain to model the state transition, that is 
     \begin{equation}
p(\overline {x}_n|\overline {x}_{n-1})=\prod_{i=1}^{\overline J}\prod_{j=1}^{\overline J} {\overline A_{i,j}^{l(\overline {x}_{n}=j,\overline {x}_{n-1}=i)}},
  \label{eq10}
\end{equation}
     \begin{equation}
p(\overline {x}_1|\overline {x}_0)=p(\overline {x}_1)=\prod_{j=1}^{\overline J}{\overline {\pi}_{j}^{l(\overline {x}_{1}=j)}},
  \label{eq11}
\end{equation}
\newline
where $l(\cdot)$ {denotes} an indicator function, which is 1 when the logic expression in the parentheses is true and 0 otherwise. In addition, $\overline A_{i,j}$ and $\overline {\pi}_{j}$ denote the transition probability from  the state $i$ to the state $j$ and the initial probability for the first frame's state $\overline{x}_1$ being state $j$, respectively. Collecting all the initial and transition probabilities, we can write them into matrix forms, i.e., $\overline{\bm{\pi}}=[\overline{\pi}_1, \cdots,\overline{\pi}_j, \cdots, \overline{\pi}_{\overline{J}}] ^T$ and $\overline{\mathbf{A}}$ with $\overline A_{i,j}$ being the element at the $i^\mathrm{th}$ row and $j^\mathrm{th}$ column.  Therefore, the modeling parameters of the HMM can be expressed as $\bm{\Phi}_\text{hmm}=\{\overline{\mathbf{A}}, \overline{\bm{\pi}}, \overline{J}\}$. The modeling parameters $\overline{\mathbf{A}}$ and $\overline{\bm{\pi}}$ with a predefined $\overline{J}$ can be trained through the EM algorithm shown in the next section.  In the experiments, we investigate the impact of the total number of states $\overline{J}$.
\newline
\textbf{The state-conditioned likelihood function}  $p({\bf s}_n|\overline {x}_n)$: Next, we  present the the proposed state-conditioned likelihood function.  Motived by the good speech enhancement performance, the computational efficient MU rule, and the equivalence between the gradient descent derivation and the EM algorithm for the KL divergence-based NMF, we propose to use the statistical model \eqref{eq7} and \eqref{eq8} to build the state-conditioned likelihood function, that is 
 \begin{equation}
   {\bf s}_n = \sum_{k=1}^{\overline K} {{\overline{\bf{c}}_n}}(k),
  \label{eq12}
\end{equation}
\begin{equation}
   {p(\overline{\bf c}_n(k)|\overline {x}_n)} = \prod_{f=1}^F  \mathcal{PO} ({\overline c}_{f,n}(k);\overline {W}_{f,k}^{\overline {x}_n}\overline {H}_{k,n}^{\overline {x}_n}),
  \label{eq13}
\end{equation}
where \(\overline{K}\) is the number of basis vectors, $\overline{\bf c}_n(k)$ contains the hidden variables, $\overline {W}_{k,n}^{\overline {x}_n}$ and $\overline {H}_{k,n}^{\overline {x}_n}$ correspond to the elements of the basis and activation matrices. Writing $\overline{\mathbf{c}}_n=[\overline{\mathbf{c}}_{n}(1)^T,\overline{\mathbf{c}}_{n}(2)^T, \cdots, \overline{\mathbf{c}}_{n}(\overline{K})^T]^T$ and integrating $\overline{\mathbf{c}}_n$ out, the state conditioned likelihood function can be written as 
  \begin{equation}
   \begin{aligned}
   &{p({\bf s}_n| \overline {x}_n)} = \int {p({\bf s}_n|{\overline{\bf{c}}_n})}{p(\overline{\bf{c}}_n|{\overline{{x}}_n})}\, d{\overline{\bf{c}}_n} \\
   & = \prod_{f=1}^F \mathcal{PO} (|S(f,n)|;\sum_{k=1}^{\overline K} \overline {W}_{f,k}^{\overline {x}_n}\overline {H}_{k,n}^{\overline {x}_n}),
  \end{aligned}
  \label{eq14}
\end{equation}
where we use the superposition property of the Poisson random variable \cite{cemgil2009bayesian}. Collecting the unknown parameters $\{\overline {W}_{f,k}^{\overline {x}_n}\}$ and $\{\overline {H}_{k,n}^{\overline {x}_n}\}$, we can write them into matrix forms, i.e., $\{\overline{\mathbf{W}}^j\}$ and $\{\overline{\mathbf{H}}^j\}$. Therefore, unlike the traditional NMF using only one basis matrix, the proposed model have $\overline{J}$ basis matrices to be trained.  Each basis matrix is intended to capture a specific feature  (e.g., phoneme) of the speech signal. The modeling parameters of the proposed state-conditioned likelihood function can be expressed as $\bm{\Phi}_\text{like}=\{ \{\overline{\mathbf{W}}^j\}, \{\overline{\mathbf{H}}^j\},  \overline{K}, \overline{J} \}$. The modeling parameters $\{\overline{\mathbf{W}}^j\}$ and $\{\overline{\mathbf{H}}^j\}$ with predefined $\overline{J}$ and $\overline{K}$  can be trained through the EM algorithm shown in the next section.  In the experiments, we investigate the impact of the number of basis vectors $\overline{K}$ and $\overline{J}$.
It will also be shown that a multiplicative update rule can be derived for the basis and activation matrices update of the proposed state-conditioned likelihood function. 

To summarize,  five types of parameters in the parameter set $\boldsymbol \Phi$=$\bm{\Phi}_\text{hmm}\cup \bm{\Phi}_\text{like}$ can be identified. They are the transition matrix \(\overline{\bf{A}}\), initial state probabilities in $\overline{\boldsymbol{\pi}}$, basis matrices of different states $\{\overline{\bf{W}}^j\}$, activation matrices of different states $\{{\overline{\bf{H}}}^j\}$, and modeling parameters $\overline K$ and $\overline J$. In this paper, the  modeling parameters $\overline K$ and $\overline J$ are predefined, the activation matrices $\{\overline{\bf{H}}^j\}$ is estimated by online speech enhancement and the other three types of parameters are obtained using offline learning.
}

\subsection{Noisy Speech Model}

Based on the proposed clean speech and noise signal models, (\ref{eq1}) and (\ref{eq2}), the noisy speech model can be defined. We assume that there are a total of $\ddot {J}$ hidden states for the noise and the hidden state of the noise is $\ddot {x}_n  (\ddot {x}_n\in {\{1,2,\cdots,\ddot J\}})$. The $\ddot{\boldsymbol{\pi}}$ and \(\ddot{\bf{A}}\) correspond to the initial state probability and transition probability matrix of the noise. Thus, there are a total of $\overline{J} \times \ddot J$ hidden states for {the} noisy speech. Each composite state consists of {a pair of states of} clean speech ${\overline {x}_n}$ and noise $\ddot {x}_n$. Thus, if we list the state space for noisy signal, we have $(\overline {x}_n = 1, \ddot {x}_n = 1),(\overline {x}_n = 1, \ddot {x}_n = 2),\cdots,(\overline {x}_n = 1, \ddot {x}_n = \ddot J);(\overline {x}_n = 2, \ddot {x}_n = 1),(\overline {x}_n = 2, \ddot {x}_n = 2),\cdots,(\overline {x}_n = 2, \ddot {x}_n = \ddot J);\cdots; (\overline {x}_n = \overline J, \ddot {x}_n =1),(\overline {x}_n = \overline J, \ddot {x}_n = 2),\cdots,(\overline {x}_n = \overline J, \ddot {x}_n = \ddot J)$. Moreover, the initial state and transition probabilities matrices of the noisy speech can be expressed as $\overline{\boldsymbol{\pi}} \otimes \ddot{\boldsymbol{\pi}}$ and $\overline{\bf{A}} \otimes \ddot{\bf{A}}$, where the ${\otimes}$ denotes the Kronecker product. Finally, the {state} conditioned likelihood function of {the} noisy speech can be written as follows{:}
  \begin{equation}
  \begin{aligned}
   &{p({\bf y}_n | \overline {x}_n, \ddot {x}_n) } = \\ 
   & \prod_{f=1}^F \mathcal{PO}{(} |(Y(f,n)|;\sum_{k=1}^{\overline K} \overline {W}_{f,k}^{\overline {x}_n}\overline {H}_{k,n}^{\overline {x}_n}+\sum_{k=1}^{\ddot K} \ddot {W}_{f,k}^{\ddot {x}_n}\ddot {H}_{k,n}^{\ddot {x}_n}),
   \end{aligned}
  \label{eq15}
\end{equation}
where $\ddot K$, $\{\ddot {W}_{f,k}^{\ddot {x}_n}\}$, and $\{\ddot {H}_{f,k}^{\ddot {x}_n}\}$ represent the number of basis {vectors}, elements of the basis matrices {and the activation matrices for the noise}, respectively. {We can write $\{\ddot {W}_{f,k}^{\ddot {x}_n}\}$ and $\{\ddot {H}_{k,n}^{\ddot {x}_n}\}$ into matrix forms $\{{\ddot{\bf{W}}}^j\}$ and $\{{\ddot{\bf{H}}}^j\}$.} {Note that, we also used the superposition property of Poisson random variables to obtain \eqref{eq15}.}

\section{Offline NMF-HMM-based parameter learning}

In the offline training stage, the {objective} is to find the parameter set $\boldsymbol \Phi$ that maximizes the likelihood function {(\ref{eq9})}. In general, the EM algorithm \cite{baum1972inequality} can be used to address this problem. {Since we used the same model for the speech and the noise, we here use the clean speech as the example to illustrate the offline parameter learning process.} First, we define the complete data set $({\bf S}_N, {\bf{\overline{x}}}_N, {\bf{\overline{C}}}_N)$, where ${\bf{\overline{C}}}_N=[\overline{\bf c}_1,\overline{\bf c}_2,\cdots,\overline{\bf c}_N]$. Thus, using the conditional independence property, the complete data likelihood function can be written as  
  \begin{equation}
   \begin{aligned}
  &{p({\bf{S}}_N, {\bf{\overline{x}}}_N, {\bf{\overline{C}}}_N)}=\prod_{n=1}^N p({\bf s}_n|\overline{\bf c}_n)p(\overline{\bf c}_n|\overline {x}_n)p(\overline {x}_n|\overline {x}_{n-1}).
     \end{aligned}
  \label{eq16}
\end{equation}
{Next, we show how the parameter set can be obtained iteratively using the EM algorithm. Moreover, we propose a speeding up strategy to lower the computational and memory complexities. The traditional MU update algorithm for the KL divergence-based NMF can be seen as a special case of the proposed algorithm.}  
\newline
\noindent{\bf Expectation step:} We first calculate the {posterior state} probability and {the} joint posterior probability, which can be written as 
 \begin{equation}
   q(\overline {x}_{n})=p(\overline {x}_{n}|{\bf{S}}_N;{\boldsymbol \Phi}^{i-1}),
  \label{eq17}
\end{equation}
 \begin{equation}
   q(\overline {x}_{n},\overline {x}_{n-1})=p(\overline {x}_{n},\overline {x}_{n-1}|{\bf{S}}_N;{\boldsymbol \Phi}^{i-1}),
  \label{eq18}
\end{equation}
where $i$ is the iteration number. The calculation of  (\ref{eq17}) and  (\ref{eq18}) can be performed using the forward-backward algorithm \cite{baum1972inequality}. {Apart from this, we also} need to evaluate the posterior expectation ${\mathbb E}_{\overline{\bf c}_n|{\bf{S}}_N,{\overline {x}_{n}};{{\boldsymbol \Phi}}^{i-1}}(\overline{\bf c}_n)$, which will be used in the maximization-step. By using {the} Bayes rule and the conditional independence property of the proposed {model}, we {have}
 \begin{equation}
   q(\overline{\bf c}_n|\overline {x}_{n})= p({\overline{\bf c}_n|{\bf{S}}_N,{\overline {x}_{n}};{{\boldsymbol \Phi}}^{i-1}}) = \cfrac{{p(\bf s}_n|{\overline{\bf{c}}_n}){p(\overline{\bf{c}}_n|{\overline{{x}}_n})}}{p({\bf{S}}_N,{\overline {x}_{n})}}{.}
  \label{eq19}
\end{equation}
Combining (\ref{eq12}) and (\ref{eq13}) and following the derivation in {\cite{cemgil2009bayesian}}, we have 

 \begin{equation}
 \begin{aligned}
   &q(\overline{\bf c}_n|\overline {x}_{n})= \\
   &\prod_{f=1}^F {\mathcal{M}} ({\overline c}_{f,n}(1),\cdots,{\overline c}_{f,n}({\overline{K}});|S(f,n)|,p_{f,n}^{{\overline x}_{n}}(1),\cdots,p_{f,n}^{{\overline x}_{n}}{(\overline{K})}),
   \end{aligned}
  \label{eq20}
\end{equation}
where ${\mathcal{M}} (\cdot)$ {denotes} the multinomial distribution {and}
 \begin{equation}
p_{f,n}^{{\overline x}_{n}}(k)=\cfrac{\overline {W}_{f,k}^{\overline {x}_n}\overline {H}_{k,n}^{\overline {x}_n}}{\sum_{l=1}^{\overline{K}}\overline {W}_{f,l}^{\overline {x}_n}\overline {H}_{l,n}^{\overline {x}_n}}. 
  \label{eq21}
\end{equation}
{Using the properties of the multinomial distribution, the mean can be written as}
 \begin{equation}
{\mathbb E}({\overline c}_{f,n}({k})|{\bf{S}}_N,{\overline {x}_{n}})=|S(f,n)|\cfrac{\overline {W}_{f,k}^{\overline {x}_n}\overline {H}_{k,n}^{\overline {x}_n}}{\sum_{l=1}^K\overline {W}_{f,l}^{\overline {x}_n}\overline {H}_{l,n}^{\overline {x}_n}}. 
  \label{eq22}
\end{equation}

\noindent{\bf Maximization step:} In this step, our objective is to find parameters to maximize the {expectation of the logarithm of the complete data likelihood,} that is,

\begin{equation}
{{\boldsymbol \Phi}}^{i}= {\mathop{\arg\max}_{\boldsymbol \Phi}}  { \mathbb E}_{{\bf{\overline{x}}}_N, {\bf{\overline{C}}}_N|{\bf{S}}_N;{{\boldsymbol \Phi}}^{i-1}}[\log {p({\bf{S}}_N, {\bf{\overline{x}}}_N, {\bf{\overline{C}}}_N)}].
  \label{eq23}
\end{equation}
{The estimators for the \(\overline{\bf{A}}\) and $\overline{\boldsymbol{\pi}}$} are the same as the traditional HMM \cite{baum1972inequality}. For completeness, the results are shown below
\begin{equation}
{\overline{\pi}}_j=\cfrac{q(\overline {x}_{1}=j)}{\sum_{o=1}^{\overline {J}} q(\overline {x}_{1}=o)},
  \label{eq24}
\end{equation}
\begin{equation}
{\overline{A}}_{o,j}=\cfrac{\sum_{n=2}^{\overline {N}}q(\overline {x}_{n}=j,\overline {x}_{n-1}=o)}{\sum_{j=1}^{\overline {J}}\sum_{n=2}^{\overline {N}} q(\overline {x}_{n}=j,\overline {x}_{n-1}=o)},
  \label{eq25}
\end{equation}
where $1\leq o,j\leq \overline {J}$. The {estimated basis and activation matrices} can be derived by {setting the derivatives of (\ref{eq23}) to zeros, and we can obtain}
{
\begin{equation}
W_{f,k}^{j}=\cfrac{\sum_{n=1}^{{N}}q(\overline {x}_{n}=j){\mathbb E}({\overline c}_{f,n}(k)|{\bf{S}}_N,{\overline {x}_{n}}=j)}{\sum_{n=1}^{{N}}q(\overline {x}_{n}=j)H_{k,n}^{j}},
  \label{eq26}
\end{equation}
\begin{equation}
H_{k,n}^{j}=\cfrac{\sum_{f=1}^{{F}}{\mathbb E}({\overline c}_{f,n}(k)|{\bf{S}}_N,{\overline {x}_{n}}=j)}{\sum_{f=1}^{{F}}W_{f,k}^{j}}.
  \label{eq27}
\end{equation}

\noindent\textbf{Speeding up strategy:} Although we can directly use the above EM algorithm to update the parameter set, saving the conditional expectation of ${\overline c}_{f,n}({k})$ in \eqref{eq22} requires a lot of memory.  Like \cite{cemgil2009bayesian}, we substitute (\ref{eq22}) into (\ref{eq26}) and (\ref{eq27}), and we can obtain:
\begin{equation}
W_{f,k}^{j} \leftarrow \cfrac{\displaystyle{\sum_{n=1}^{{N}}q(\overline {x}_{n}=j)\cfrac{|S(f,n)|\overline {H}_{k,n}^{j}}{\sum_{l=1}^{\overline{K}}\overline {W}_{f,l}^{j}\overline {H}_{l,n}^{j}}}}{\sum_{n=1}^{{N}}q(\overline {x}_{n}=j)H_{k,n}^{j}},
  \label{eq28}
\end{equation}
\begin{equation}
H_{k,n}^{j} \leftarrow \cfrac{\displaystyle{\sum_{f=1}^{{F}}\cfrac{\overline {W}_{f,k}^{j}|S(f,n)|}{\sum_{l=1}^K\overline {W}_{f,l}^{j}\overline {H}_{l,n}^{j}}}}{\sum_{f=1}^{{F}}H_{k,n}^{j}}.
  \label{eq29}
\end{equation}
We can further write (\ref{eq28}) and (\ref{eq29}) in matrix forms 
     \begin{equation}
   {\overline{\bf{W}}}^{j} \leftarrow  {\overline{\bf{W}}}^{j} \odot {\cfrac{\cfrac{{\bf{S}}_N}{{\overline{\bf{W}}}^{j}{\overline{\bf{H}}}^{j}}{{\boldsymbol\Lambda} (j)}({\overline{\bf{H}}}^{j})^\mathit{T}}{{\bf{1}}({\overline{\bf{H}}}^{j})^\mathit{T}}},
  \label{eq30}
\end{equation}
     \begin{equation}
   {\overline{\bf{H}}}^{j} \leftarrow  {\overline{\bf{H}}}^{j} \odot {\cfrac{({{\overline{\bf{W}}}^{j}})^\mathit{T}\cfrac{{\bf{S}}_N}{{\overline{\bf{W}}}^{j}{\overline{\bf{H}}}^{j}}}{({\overline{\bf{W}}}^{j})^\mathit{T}\bf{1}}},
  \label{eq31}
\end{equation}
where ${\boldsymbol\Lambda} (j)={\rm diag}(q(\overline {x}_1=j),q(\overline {x}_2=j),\cdots,q(\overline {x}_N=j))$. Using the proposed speeding up strategy, the computing and saving of the conditional expectation of ${\overline c}_{f,n}({k})$ in \eqref{eq22} is not required. Moreover, the multiplicative update rules for  the basis and activation matrices can be obtained, leading to fast computing. Comparing the update rules of the proposed method (\ref{eq30}), (\ref{eq31}) and the traditional NMF-based method (\ref{eq5}), (\ref{eq6}), the difference is that the basis vectors update rule (\ref{eq30}) for the proposed method takes the posterior state information ${\boldsymbol\Lambda} (j)$ into account. In fact, if the number of state is set to 1 (i.e., $\overline{J} = 1$), the proposed training method is identical to the traditional KL divergence-based NMF approach. The entire flow of the offline parameter learning is shown in Algorithm 1. Note that, for stability reasons, each column of ${\overline{\bf{W}}}^{j}$ is normalized to have unit norm during training. }

\begin{table}[t]
  
  \label{table 1}
  \centering
  \begin{tabular}{l}
    \toprule
    \textbf{Algorithm 1:} Offline NMF-HMM-based parameter learning \\
    \midrule
    1: Randomly initiate ${\overline{\bf{W}}}^{j} $ and ${\overline{\bf{H}}}^{j},  \overline j \in \{1,2,\cdots,\overline J\}$\\
    2: \textbf{for $i=1,2,3,\cdots,I$} \textbf {do}          \\
    { }{ }{ }{ }{ } \textbf {Expectation step:}                           \\
    3:{ }{ }{ }{ }Calculate ${p({\bf s}_n| \overline {x}_n)}$, $1\leq n \leq  {N}$ based on (\ref{eq14})                                          \\
    4:{ }{ }{ }{ }Obtain (\ref{eq17}) and (\ref{eq18}) using the forward-backward algorithm \cite{baum1972inequality}                 \\
    { }{ }{ }{ }{ } \textbf {Maximization step:}                                \\           5:{}{ }{ }{ }{ }Re-estimate  $\overline{\boldsymbol{\pi}}$ and $\overline{\bf{A}}$ based on (\ref{eq24}) and (\ref{eq25})                        \\
    6:{}{ }{ }{ }{ }Re-estimate ${\overline{\bf{W}}}^{j} $ and ${\overline{\bf{H}}}^{j}$ based on (\ref{eq30}) and (\ref{eq31})          \\
    7: \textbf{end for}                                   \\
    \bottomrule
  \end{tabular}
\end{table}

{\section{Online speech enhancement using the MMSE estimator}}
\subsection{MMSE Estimator for the NMF-HMM}
{In this section, we propose a MMSE-based online speech enhancement algorithm for the proposed NMF-HMM model. The objective is to obtain the MMSE estimate of the desired clean speech signal from noisy observation, i.e.,}

\begin{equation}
{{\hat {\bf{s}}}}_n={\mathbb E}_{{\bf s}_n|{\bf{Y}}_n}({\bf s}_n)= \int {{\bf s}_n}p({{\bf s}_n|{\bf{Y}}_n})\, d{{\bf s}_n},
  \label{eq32}
\end{equation}
In (\ref{eq32}), the posterior probability $p({{\bf s}_n|{\bf{Y}}_n})$ can be derived as 
  \begin{equation}
   \begin{aligned}
   &p({{\bf s}_n|{\bf{Y}}_n}) = \cfrac{p({\bf{s}}_n,{\bf{y}}_n|{\bf{Y}}_{n-1})}{p({\bf{y}}_n|{\bf{Y}}_{n-1})} \\
   &= \cfrac{\sum_{\overline {x}_n, \ddot {x}_n} p({\bf{s}}_n,{\bf{y}}_n|{\overline {x}_n, \ddot {x}_n})p({\overline {x}_n, \ddot {x}_n}|{\bf{Y}}_{n-1})}{p({\bf{y}}_n|{\bf{Y}}_{n-1})},
  \end{aligned}
  \label{eq33}
\end{equation}
where we use the conditional independence property of the HMM. The term $p({\overline {x}_n, \ddot {x}_n}|{\bf{Y}}_{n-1})$ in (\ref{eq33}) can be expressed as 
  \begin{equation}
   \begin{aligned}
   &p({\overline {x}_n, \ddot {x}_n}|{\bf{Y}}_{n-1}) \\
   &= \sum_{\overline {x}_{n-1}, \ddot {x}_{n-1}} p({\overline {x}_n, \ddot {x}_n}|{\overline {x}_{n-1}, \ddot {x}_{n-1}})p({\overline {x}_{n-1}, \ddot {x}_{n-1}}|{\bf{Y}}_{n-1}),
  \end{aligned}
  \label{eq34}
\end{equation}
where the first term after the summation is the state transition probability for noisy signal and the second term is the forward probability that can be acquired using the well-known forward algorithm \cite{baum1972inequality}. By applying the Bayes rule, the term $p({\bf{s}}_n,{\bf{y}}_n|{\overline {x}_n, \ddot {x}_n})$ in (\ref{eq33}) can be further written as
\begin{equation}
p({\bf{s}}_n,{\bf{y}}_n|{\overline {x}_n, \ddot {x}_n}) = p({\bf{s}}_n|{\bf{y}}_n,{\overline {x}_n, \ddot {x}_n})p({\bf{y}}_n|{\overline {x}_n, \ddot {x}_n}).
  \label{eq35}
\end{equation}
Substituting (\ref{eq35}) to (\ref{eq33}), the posterior probability can be re-written as 
\begin{equation}
p({{\bf s}_n|{\bf{Y}}_{n}}) = \sum_{\overline {x}_{n-1},\ddot {x}_{n-1}} \omega_{\overline {x}_n, \ddot {x}_n}p({\bf{s}}_n|{\bf{y}}_n,{\overline {x}_n, \ddot {x}_n}),
  \label{eq36}
\end{equation}
where the weight $0 \leq \omega_{\overline {x}_n, \ddot {x}_n} \leq 1$ {is defined as}
\begin{equation}
{\omega_{\overline {x}_n, \ddot {x}_n}}=\cfrac{{p({\bf y}_n | \overline {x}_n, \ddot {x}_n) }{p(\overline {x}_n, \ddot {x}_n | {\bf Y}_{n-1}})}{\sum_{\overline {x}_n, \ddot {x}_n}{p({\bf y}_n | \overline {x}_n, \ddot {x}_n) }{p(\overline {x}_n, \ddot {x}_n | {\bf Y}_{n-1}}) }.
  \label{eq37}
\end{equation}
Thus, by combining (\ref{eq32}) and (\ref{eq36}), the proposed HMM-based MMSE estimator can be expressed as 
\begin{equation}
{{\hat {\bf{s}}}}_n = \sum_{\overline {x}_{n},\ddot {x}_{n}} \omega_{\overline {x}_n, \ddot {x}_n} \int {{\bf s}_n}p({\bf{s}}_n|{\bf{y}}_n,{\overline {x}_n, \ddot {x}_n})\, d{{\bf s}_n}.
  \label{eq38}
\end{equation}
Instead of obtaining the {posterior} probability density function (PDF) $p({\bf{s}}_n|{\bf{y}}_n,{\overline {x}_n, \ddot {x}_n})$ directly, we derive the formula for the joint posterior PDF of the clean speech and noise first, that is, 
\begin{equation}
   \begin{aligned}
   &p({{\bf s}_n,{\bf m}_n|{\bf y}_n},{\overline {x}_n, \ddot {x}_n}) \\
   &= \cfrac{p({{\bf y}_n|{\bf s}_n,{\bf m}_n})p({{\bf s}_n,{\bf m}_n|}{\overline {x}_n, \ddot {x}_n})}{p({{\bf y}_n}|{\overline {x}_n, \ddot {x}_n})} \\
   &= \cfrac{p({{\bf y}_n|{\bf s}_n,{\bf m}_n})p({{\bf s}_n|}{\overline {x}_n})p({{\bf m}_n|}{\ddot {x}_n})}{p({{\bf y}_n}|{\overline {x}_n, \ddot {x}_n})}.
  \end{aligned}
  \label{eq39}
\end{equation}
By using (\ref{eq1}), we can express the likelihood function $p({{\bf y}_n|{\bf s}_n,{\bf m}_n})$ as $p({{\bf y}_n|{\bf s}_n,{\bf m}_n})= \delta ({{\bf y}_n-{\bf s}_n-{\bf m}_n})$, where $\delta (\cdot)$ denotes the Dirac delta function, which is defined by $\delta (0)=+\infty$ and $\delta (x)=0$ when $x\ne0$. Furthermore, there is $\int_{-\infty}^{+\infty}\delta (x)\, dx=1$. The prior probability $p({{\bf s}_n|}{\overline {x}_n})$ and $p({{\bf m}_n|}{\ddot {x}_n})$ can be estimated by using (\ref{eq14}). Following the derivation in \cite{cemgil2009bayesian}, we can verify that the joint posterior PDF can be expressed in terms of the multinomial distribution as 
 \begin{equation}
 \begin{aligned}
   &p({{\bf s}_n,{\bf m}_n|{\bf y}_n},{\overline {x}_n, \ddot {x}_n})= \\
   &\prod_{f=1}^F {\mathcal{M}} (|S(f,n)|,|M(f,n)|;|Y(f,n)|,p_{f,n}(\overline {x}_n, \ddot {x}_n),q_{f,n}(\overline {x}_n, \ddot {x}_n)),
   \end{aligned}
  \label{eq40}
\end{equation}
where $p_{f,n}(\overline {x}_n, \ddot {x}_n)$ and $q_{f,n}(\overline {x}_n, \ddot {x}_n)$ {are defined as} 
 \begin{equation}
 \begin{aligned}
   &p_{f,n}(\overline {x}_n, \ddot {x}_n)= \\
   &\cfrac{\sum_{k=1}^{\overline K}\overline {W}_{f,k}^{\overline {x}_n}\overline {H}_{k,n}^{\overline {x}_n}}{\sum_{k=1}^{\overline K} \overline {W}_{f,k}^{\overline {x}_n}\overline {H}_{k,n}^{\overline {x}_n}+\sum_{k=1}^{\ddot K} \ddot {W}_{f,k}^{\ddot {x}_n}\ddot {H}_{k,n}^{\ddot {x}_n}},
   \end{aligned}
  \label{eq41} 
\end{equation}
where $q_{f,n}(\overline {x}_n, \ddot {x}_n)=1-p_{f,n}(\overline {x}_n, \ddot {x}_n)$. Therefore, the integral term in (\ref{eq38}) can be expressed as 
\begin{equation}
 \begin{aligned}
   &\int {{\bf s}_n}p({\bf{s}}_n|{\bf{y}}_n,{\overline {x}_n, \ddot {x}_n})\, d{{\bf s}_n} \\
   &=\int {{\bf s}_n} \int {p({\bf{s}}_n,{\bf{m}}_n|{\bf{y}}_n,{\overline {x}_n, \ddot {x}_n})}\, d{ {\bf m}_n}\, d{ {\bf s}_n} \\
   &={{\bf{y}}_n} \odot {{\bf p}_n}({\overline {x}_n, \ddot {x}_n}),
   \end{aligned}
  \label{eq42} 
\end{equation}
where ${{\bf p}_n}({\overline {x}_n, \ddot {x}_n}) = [p_{1,n}(\overline {x}_n, \ddot {x}_n),\cdots,p_{F,n}(\overline {x}_n, \ddot {x}_n)]^T $, and we {used} the marginal mean property of the multinomial distribution. Combing (\ref{eq38}) and (\ref{eq42}), the MMSE estimator can be expressed as:
\begin{equation}
{{\hat {\bf{s}}}}_n =  {{\bf y}_n} \odot {{\bf g}_n},
  \label{eq43}
\end{equation}
\begin{equation}
{{\bf g}_n} =  {\sum_{\overline {x}_n, \ddot {x}_n} \omega_{\overline {x}_n, \ddot {x}_n}{{\bf p}_n}({\overline {x}_n, \ddot {x}_n}}),
  \label{eq44}
\end{equation}
where ${\bf g}_n$ can be viewed as the spectral gain vector for the proposed model. Comparing the proposed gain vector ${\bf g}_n$ and the traditional NMF-based gain vector \cite{grais2011single}, we find that the proposed gain vector is a weighted sum of each state’s gain, {which is in the Wiener filtering form as the traditional NMF gain \eqref{nmfgain}. }
\subsection{Online Estimation of Activation Matrices}
After obtaining the trained basis matrices $\overline {W}_{f,k}^{\overline {x}_n}$ and $\ddot {W}_{f,k}^{\ddot {x}_n}$ for both {the} clean speech and noise in the training stage, we need to {obtain the online estimates of  the activation parameters} $\overline {H}_{f,k}^{\overline {x}_n}$ and $\ddot {H}_{f,k}^{\ddot {x}_n}$ to acquire the gain in (\ref{eq43}) and (\ref{eq44}). The activation matrices are estimated by maximizing the logarithm of the state-conditioned likelihood function (\ref{eq15}), which is equivalent to 
     \begin{equation}
  {\bf{h}}_n ({\overline {x}_n, \ddot {x}_n}) = {\mathop{\arg\min}_{{\bf{h}}_n}}  \ \ { \rm KL}({{\bf y}}_n|[{\overline{\bf{W}}}^{\overline {x}_n},{\ddot{\bf{W}}}^{\ddot {x}_n}]{\bf{h}}_n),
  \label{eq45}
\end{equation}
     \begin{equation}
  {\bf{h}}_n ({\overline {x}_n, \ddot {x}_n}) = [{\overline{\bf{h}}_n}({\overline {x}_n, \ddot {x}_n})^T,{\ddot{\bf{h}}_n}({\overline {x}_n, \ddot {x}_n})^T]^T,
  \label{eq46}
\end{equation}
where the clean and noise activation matrices for the state $({\overline {x}_n, \ddot {x}_n})$ are defined as ${\overline{\bf{h}}_n} ({\overline {x}_n, \ddot {x}_n})= [\overline {H}_{1,n}^{\overline {x}_n},\overline {H}_{2,n}^{\overline {x}_n},\cdots,{H}_{\overline {K},n}^{\overline {x}_n}]^T$ and ${\ddot{\bf{h}}_n} ({\overline {x}_n, \ddot {x}_n})= [\ddot {H}_{1,n}^{\ddot {x}_n},\ddot {H}_{2,n}^{\ddot {x}_n},\cdots,{H}_{\ddot {K},n}^{\ddot {x}_n}]^T$. The {activation matrix  (\ref{eq46}) can be obtained iteratively} by using the multiplicative update rule in equation (\ref{eq6}). {Note that, to obtain the activation matrices for different states, parallel computing can be used to reduce the time complexity. It can be readily shown that when $\overline{J} = \ddot{J}=1$, the gain vectors for the proposed algorithm \eqref{eq44} and the standard NMF \eqref{nmfgain} are identical, that is $\mathbf{g}_n=\mathbf{g}_n^{\text{NMF}}$. The entire flow of  proposed MMSE-based online speech enhancement algorithm is illustrated by Algorithm 2.}

\begin{table}[t]
  
  \label{table 2}
  \centering
  \begin{tabular}{l}
    \toprule
    \textbf{Algorithm 2:} MMSE-based online speech enhancement \\
    \midrule
    1: Input magnitude spectrum: ${\bf{Y}}_n$\\
    2: Initiate $\overline{\boldsymbol{\pi}} \otimes \ddot{\boldsymbol{\pi}}$ and $\overline{\bf{A}} \otimes \ddot{\bf{A}}$\\
    3: \textbf{for $n=1,2,3,\cdots,N$} \textbf {do}          \\
    4:{ }{ }{ }{ }Initiate ${\bf{h}}_n ({\overline {x}_n, \ddot {x}_n})$   \\
    5:{ }{ }{ }{ }Based on (\ref{eq6}) and (\ref{eq46}), obtain the iterative estimation ${\bf{h}}_n ({\overline {x}_n, \ddot {x}_n})$               \\
    6:{}{ }{ }{ }{ }Calculate   $p({\bf y}_n | \overline {x}_n, \ddot {x}_n)$ based on (\ref{eq15}) \\
    7:{}{ }{ }{ }{ }Apply the forward algorithm and combine (\ref{eq34}) and (\ref{eq37}) to acquire\\
    {}{ }{ }{ }{ }{ }{ }{ }${\omega_{\overline {x}_n, \ddot {x}_n}}$ \\
    7:{}{ }{ }{ }{ }Obtain ${{\bf p}_n}({\overline {x}_n, \ddot {x}_n})$ using (\ref{eq41}) \\ 8:{}{ }{ }{ }{ }Calculate the spectral gain ${\bf g}_n$ using (\ref{eq44})   \\
    9:{}{ }{ }{ }{ }By equation(\ref{eq43}), estimate the clean speech ${{\hat {\bf{s}}}}_n$ \\
    10: \textbf{end for}                                   \\
    \bottomrule
  \end{tabular}
\end{table}

\section{Analysis of experiments and results}

{In this section, we investigate and evaluate the proposed algorithm using various experiments. First, we investigate the effect of different parameter settings for the proposed model, that is the number of states and basis vectors of clean speech and noise, respectively. Second, we compare the proposed NMF-HMM with other state-of-the-art speech enhancement methods to demonstrate the effectiveness of the proposed algorithm. In this work, the PESQ \cite{rix2001perceptual}, ranging from -0.5 to 4.5, is used to quantify the enhanced speech quality. The STOI \cite{taal2011algorithm}, ranging from 0 to 1, is used to measure the speech intelligibility.}

\subsection{Experimental data preparation}
\label{sub:dataprep}
In this study, the proposed algorithm is {evaluated using} the TIMIT \cite{garofolo1993darpa} and NOISEX-92 \cite{varga1993assessment} databases. During the training stage, all 4620 utterances from the TIMIT training database are used to train the proposed NMF-HMM model for the clean speech. {The} Babble, F16, Factory and White noise from the NOISEX-92 database are used to train the NMF-HMM model for the noise. For the experiments in Section \ref{sub:dimensionAnalysisExp}, 200 utterances from the TIMIT test set, including 1680 utterances, are randomly chosen to build the test database. Then, four types of noise are added at four different SNR levels (i.e., -5, 0, 5, and 10 dB). The noise types of the testing set are the same as the training set, but no overlap between the signals in the two sets. In total, $200\times4\times4=3200$ utterances are used for evaluation. For the experiments in Section \ref{sub:overallcompare},  apart from the aforementioned four types of noise, we further add the destroyerengine and destroyerops noise (from the NOISEX-92 database) to the test set to evaluate the performance of various speech enhancement algorithms. Note that, these two noise types are not included in the training set. In all experiments, {the sound} signals are down-sampled to 16 kHz. The frame length is set to 1024 samples (64 ms) with a frame shift of 512 samples (32 ms). The size of STFT is 1024 points with a Hanning window. Furthermore, the maximum number of iterations {is set to} 30 in the training stage and 15 in the online speech enhancement stage for the proposed NMF-HMM algorithm. 

\subsection{Number of states and basis vectors analysis}
\label{sub:dimensionAnalysisExp}
\begin{figure}[!t]
  \centering
  \includegraphics[scale=0.8]{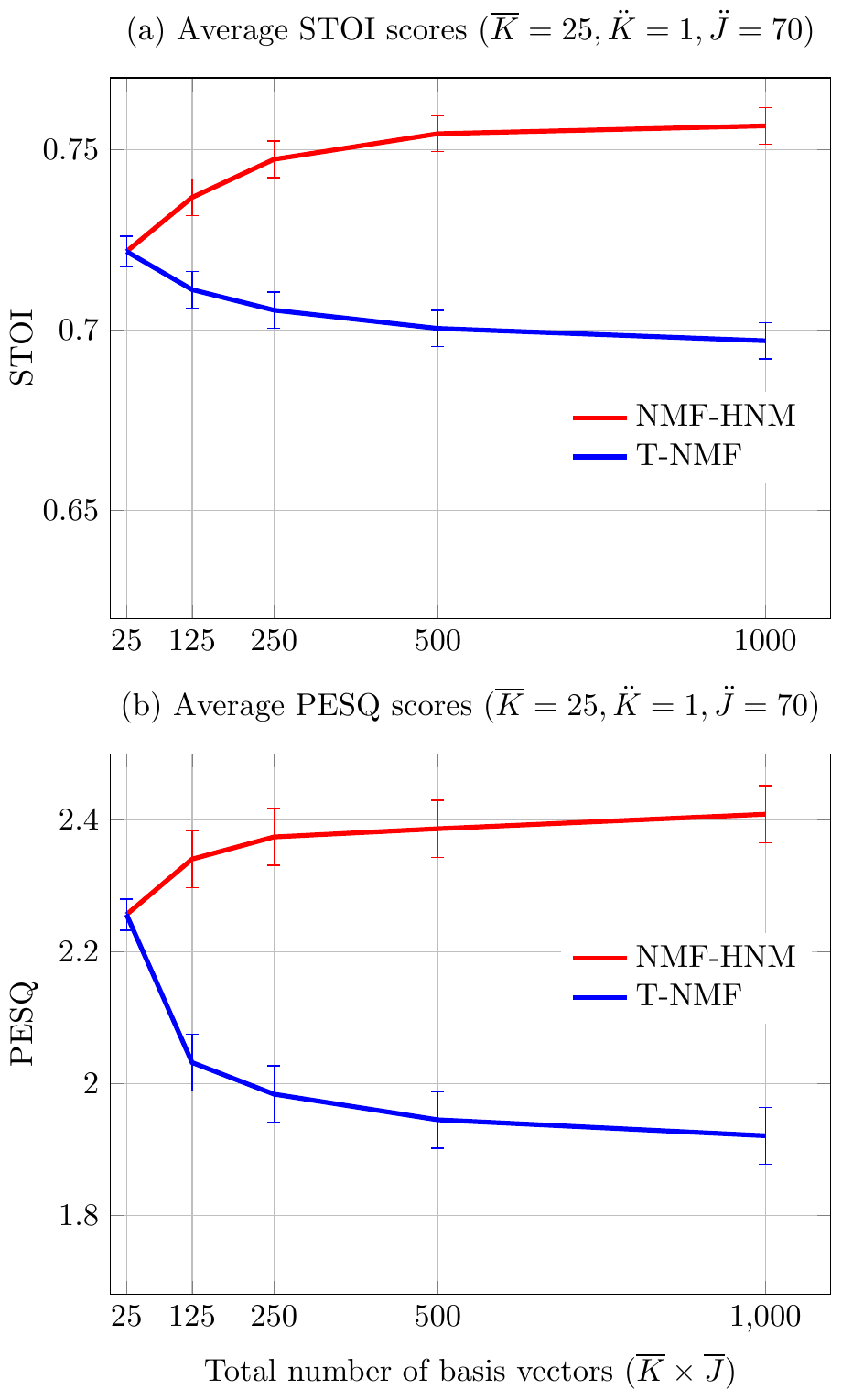}
  \caption{Performance of the NMF-HMM and T-NMF using different numbers of clean speech basis vectors.}
  \label{fig:vector state}
\end{figure}

{As explained in Sections III and IV, four parameters need to be {pre-defined} in our proposed NMF-HMM-based speech enhancement algorithm. These parameters are number of states ($\overline{J}$ and $\ddot{J}$) and basis vectors ($\overline{K}$ and $\ddot{K}$) for the clean speech and noise. In this part, we investigate the effects of these parameters in our proposed method and choose suitable parameters for the later experiments.}


First, we investigate the effect of total numbers of clean speech basis vectors ($\overline{K}\times\overline{J}$) for the NMF-HMM and T-NMF. In this experiment, we use the average STOI and PESQ scores of 3200 utterances as the performance metrics. In this experiment, the number of noise basis vectors for both the proposed NMF-HMM and T-NMF is fixed to 70 and the number of noise states for the NMF-HMM is fixed to 1. For the T-NMF, the number of clean speech basis vectors ${\overline{K}}$ is varied as 25, 125, 250, 500 and 1000. For the NMF-HMM, the $\overline{K}$ is fixed to 25 and $\overline{J}$ is varied as 1, 5, 10, 20 and 40 . This parameter setting ensures that the total number of clean speech basis vectors is the same for the T-NMF and NMF-HMM. The experimental results are shown in Figure~\ref{fig:vector state}. As can be seen, the T-NMF can achieve the best performance when $\overline{K} = 25.$ However, its performance degraded with the increasing of number of basis vectors due to overfitting. By contrast, NMF-HMM achieves higher PESQ and STOI scores with an increasing number of clean speech basis vectors by taking the temporal dynamics into account using the HMM model. Furthermore, it can be found that NMF-HMM can achieve a 5\% improvement for STOI and 0.18 improvement for PESQ than T-NMF.

\subsubsection{Number of states analysis}
{Then, we investigate the effect of the number of clean speech states (i.e., $\overline{J}$) to the proposed model. The number of noise states is set to 2 (i.e., $\ddot J = 2$) for the proposed NMF-HMM. The number of  basis vectors for the clean speech and noise is fixed to ${\overline{K}}=25$ and ${\ddot{K}}=70$, respectively. The number of clean speech states is chosen as 1, 5, 10, 20 and 40. The traditional NMF-based speech enhancement (T-NMF) method $(\ddot J = 1,\overline{J} = 1,{\overline{K}}=25,{\ddot{K}}=70)$ \cite{grais2011single} is used as the reference method. The enhancement performance is evaluated by the PESQ and STOI.}

\begin{figure}[!t]
  \centering
  \includegraphics[width=\linewidth]{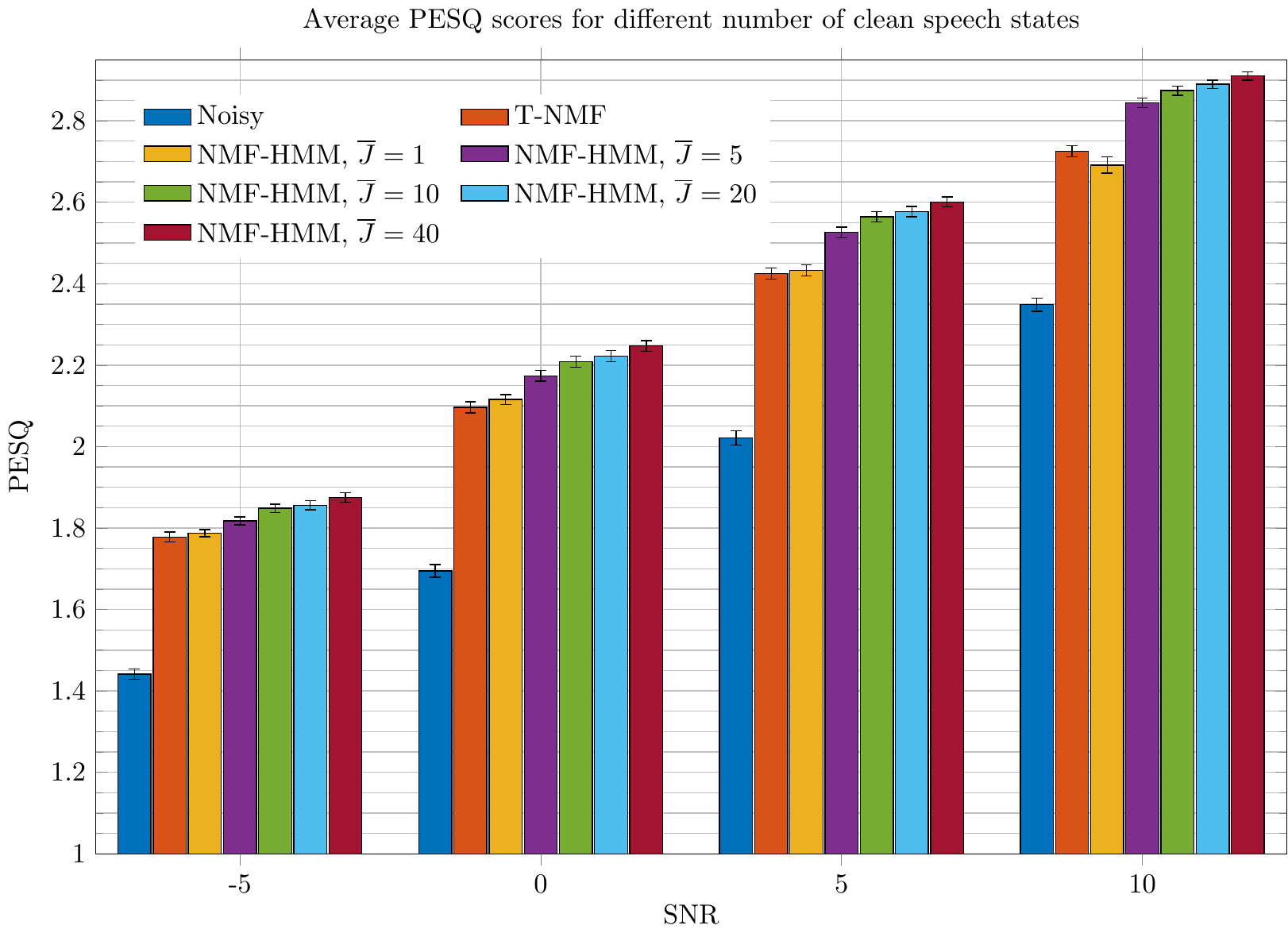}
  \caption{Average PESQ scores of proposed algorithm with different number of clean states (${\overline{K}}=25,\ddot J = 2,{\ddot{K}}=70$).}
  \label{fig:clean state}
\end{figure}
\begin{table*}[!tbp]
  \caption{Average STOI scores (\%) of different clean speech state (${\overline{K}}=25,\ddot J = 2,{\ddot{K}}=70$)}
  \label{tab:clean state}
  \centering
  \begin{tabular}{ccccc}
    \toprule
    \textbf{SNR(dB)} & \textbf{-5} & \textbf{0} & \textbf{5} & \textbf{10} \\
    \midrule
    \textbf{Noisy}& 51.59 ($\pm\,{0.50}$) &	64.26 ($\pm\, 0.52)$ & 76.30 ($\pm\, 0.47$) & 85.86 ($\pm\, 0.36)$ \\
    \midrule
    \textbf{T-NMF}& 55.85 ($\pm\,0.76$) &68.96 ($\pm\,0.73$) & 79.51 ($\pm\,0.59$) & 86.78 ($\pm\,0.40$) \\
    \midrule
    \textbf{NMF-HMM, ${\overline{J}}=1$}& 56.14 ($\pm\,0.76$) & 69.21 ($\pm\,0.72$) & 79.29 ($\pm\,0.57$) & 85.80 ($\pm\,0.39$) \\
    \midrule                   
    \textbf{NMF-HMM, ${\overline{J}}=5$}& 57.32 ($\pm\,0.80$) & 71.00 ($\pm\,0.74$) & 81.16 ($\pm\,0.58$) & 87.94 ($\pm\,0.38$) \\
    \midrule  
    \textbf{NMF-HMM, ${\overline{J}}=10$}& 58.54 ($\pm\,0.83$) & 72.35 ($\pm\,0.74$) & 82.23 ($\pm\,0.56$) & 88.80 ($\pm\,0.37$)\\
    \midrule 
    \textbf{NMF-HMM, ${\overline{J}}=20$}& 59.56 ($\pm\,0.84$) & 73.24 ($\pm\,0.74$) & 82.92 ($\pm\,0.54$) & 89.10 ($\pm\,0.36$)\\
    \midrule
    \textbf{NMF-HMM, ${\overline{J}}=40$}& \textbf{59.84 ($\pm\,0.87$)} & \textbf{73.46 ($\pm\,0.75$)} & \textbf{83.17 ($\pm\,0.55$)} & \textbf{89.33 ($\pm\,0.36$)}\\
    
    \bottomrule                             
  \end{tabular}
\end{table*}

\begin{figure}[!t]
  \centering
  \includegraphics[width=\linewidth]{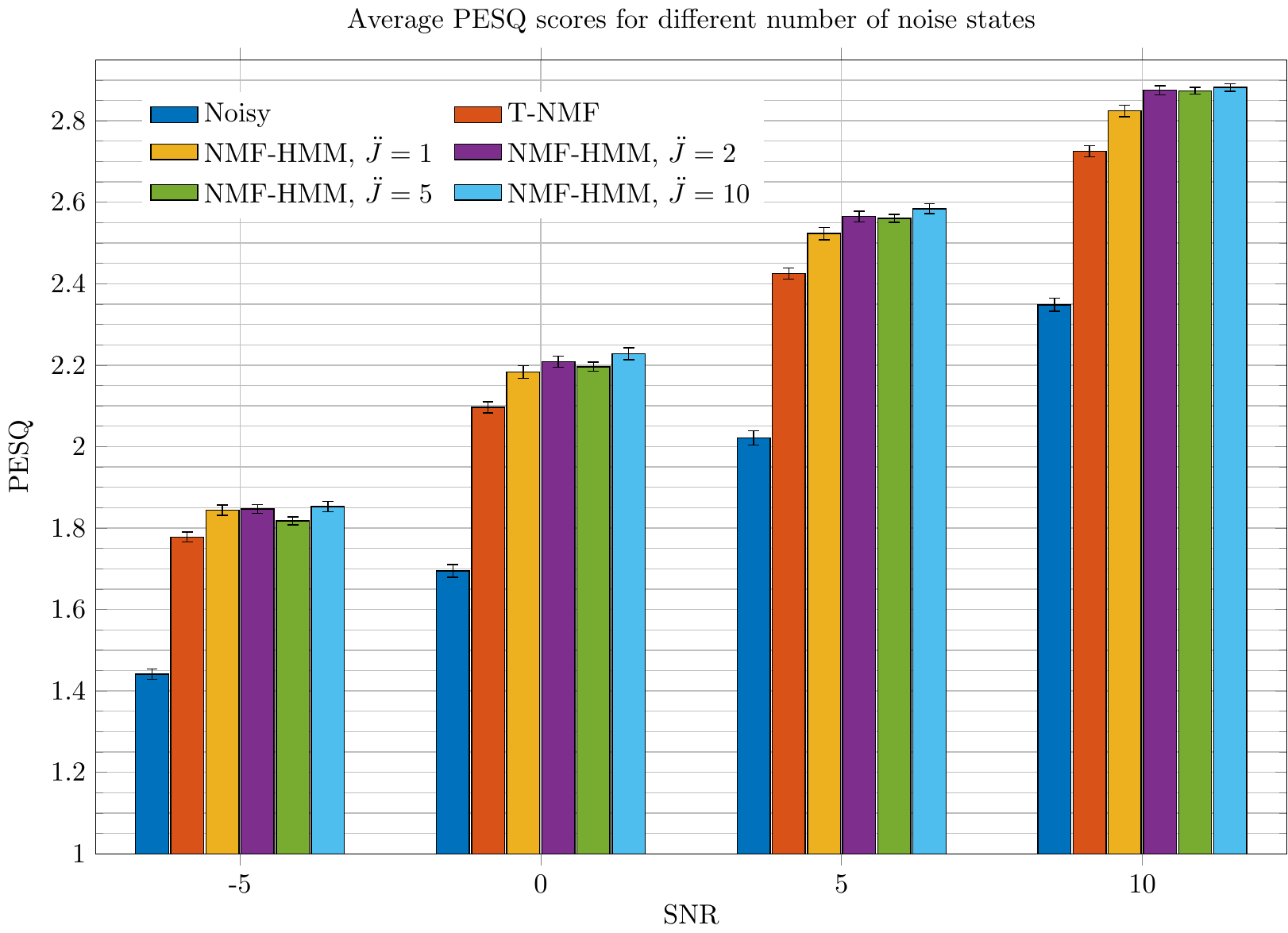}
  \caption{Average PESQ scores of proposed algorithm with different numbers of noise state ($\overline{J} = 10,{\overline{K}}=25,{\ddot{K}}=70$).}
  \label{fig:noise state}
\end{figure}

\begin{table*}[!tbp]
  \caption{Average STOI scores (\%) of different noise state ($\overline{J} = 10,{\overline{K}}=25,{\ddot{K}}=70$)}
  \label{tab:noise state}
  \centering
  \begin{tabular}{ccccc}
    \toprule
    \textbf{SNR(dB)} & \textbf{-5} & \textbf{0} & \textbf{5} & \textbf{10} \\
    \midrule
    \textbf{Noisy}& 51.59 ($\pm\,{0.50}$) &	64.26 ($\pm\, 0.52)$ & 76.30 ($\pm\, 0.47$) & 85.86 ($\pm\, 0.36)$ \\
    \midrule
    \textbf{T-NMF}& 55.85 ($\pm\,0.76$) &68.96 ($\pm\,0.73$) & 79.51 ($\pm\,0.59$) & 86.78 ($\pm\,0.40$) \\
    \midrule
    \textbf{NMF-HMM, ${\ddot{J}}=1$}& 57.71 ($\pm\,0.83$) & 71.50 ($\pm\,0.75$) & 81.79 ($\pm\,0.59$) & 88.62 ($\pm\,0.38$) \\
    \midrule                   
    \textbf{NMF-HMM, ${\ddot{J}}=2$}& 58.54 ($\pm\,0.83$) & 72.35 ($\pm\,0.74$) & 82.23 ($\pm\,0.56$) & 88.80 ($\pm\,0.37$)\\
    \midrule  
    \textbf{NMF-HMM, ${\ddot{J}}=5$}& 58.86 ($\pm\,0.82$) & 72.74 ($\pm\,0.73$) & 82.73 ($\pm\,0.55$) & 89.13 ($\pm\,0.37$)\\
    \midrule  
    \textbf{NMF-HMM, ${\ddot{J}}=10$}& \textbf{59.42 ($\pm\,0.82$)} & \textbf{73.08 ($\pm\,0.73$)} & \textbf{82.78 ($\pm\,0.55$)} & \textbf{89.20 ($\pm\,0.37$)}\\
    
    \bottomrule                             
  \end{tabular}
\end{table*}

Figure~\ref{fig:clean state} shows the average PESQ score for different number of clean speech states $\overline{J}$ in different SNRs. It can be seen that $\overline{J} =1,\ddot J = 2 $ has a slightly better PESQ score than the T-NMF ($\overline{J} =1,\ddot J = 1 $) in the low-SNR environment. In addition, with an increasing $\overline{J}$, the proposed NMF-HMM method has a consistently increasing PESQ score. In the present setting, the highest PESQ score is achieved when the $\overline{J}$ is 40. TABLE~\ref{tab:clean state} indicates the average STOI scores with 95\% confidence interval in different SNRs. As shown in TABLE~\ref{tab:clean state}, the NMF-HMM with $\overline{J} =1$ and $\ddot J = 2 $ has a similar performance to the T-NMF. Moreover, for the proposed NMF-HMM, the STOI improvement from $\overline{J} =1$ to $\overline{J} =10$ is larger than from $\overline{J} =10$ to $\overline{J} =40$. Furthermore, by taking the temporal dynamics of the speech signal into account, the proposed NMF-HMM with $\overline{J}>1$ (it also increases the total number of basis vectors) has a consistently better STOI score than the T-NMF. The proposed method with $\overline{J} = 40$ has the best performance in all SNRs.


Next, we investigate the influence of the number of noise states. The number of clean speech states is set to $\overline J = 10$. The number of noise states ${\ddot{J}}$ is chosen as 1, 2, 5 and 10. All the other settings are the same as the previous experiment. Figure ~\ref{fig:noise state} shows the average PESQ scores for different number of noise states ${\ddot{J}}$ in different SNRs. It can be seen that using a larger ${\ddot{J}}$ tends to slightly improve the PESQ score, especially in high SNR scenarios (e.g., 5-10 dB). Table~\ref{tab:noise state} illustrates the average STOI scores with 95\% confidence interval in different SNRs. In general, a larger number of noise states helps the NMF-HMM achieve higher STOI score. The highest STOI score is achieved when ${\ddot{J} = 10}$. However, the STOI improvement from increasing ${\ddot{J}}$ is limited. For example, there is no significant difference between $\ddot{J} = 2$ and $\ddot{J} = 10$. Therefore, we can conclude that the influence of number of noise states $\ddot{J}$ is less sensitive than the influence of clean speech states $\overline J$ for the proposed NMF-HMM method. This might be due to the noise being fairly stationary in this experiment and there are more changes for clean speech.

\begin{figure}[!t]
  \centering
  \includegraphics[width=\linewidth]{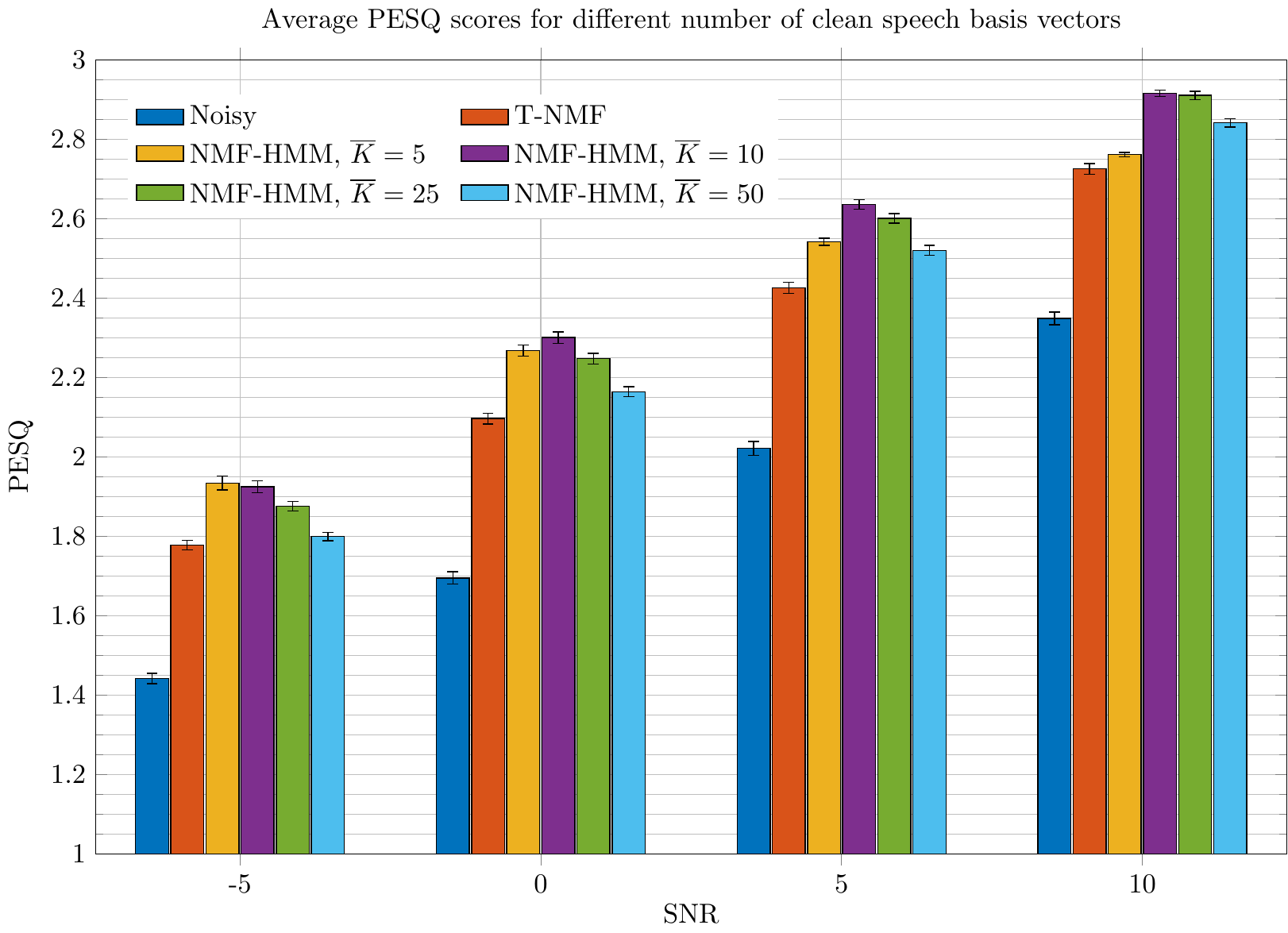}
  \caption{Average PESQ scores of different numbers of clean speech basis vectors $(\ddot J = 2,\overline{J} = 40,{\ddot{K}}=70)$.}
  \label{fig:clean dimension}
\end{figure}
\begin{figure}[!t]
  \centering
  \includegraphics[width=\linewidth]{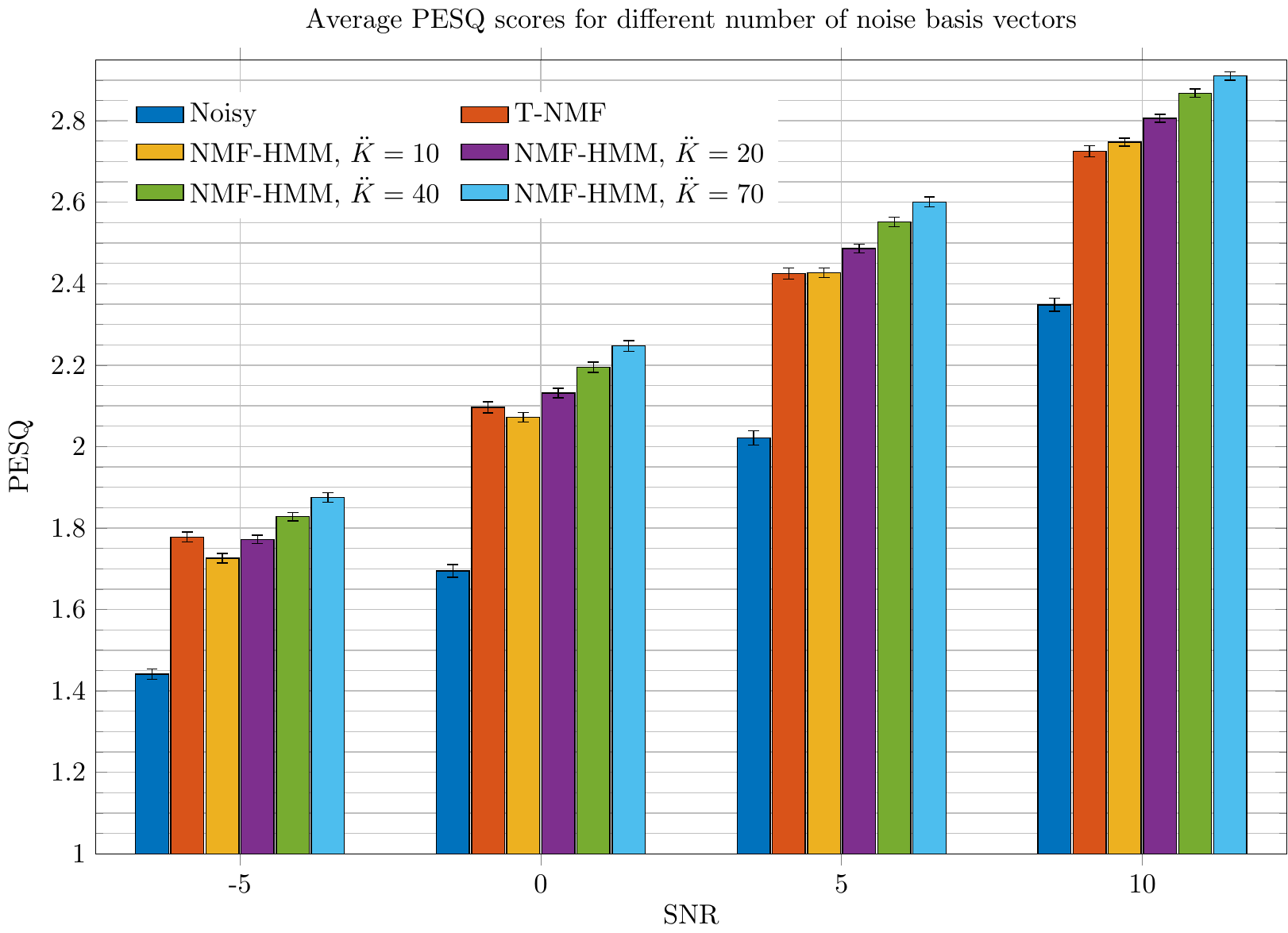}
  \caption{Average PESQ scores of different numbers of noise basis vectors $(\ddot J = 2,\overline{J} = 40,{\overline{K}}=25)$.}
  \label{fig:noise dimension}
\end{figure}

\subsubsection{Number of basis vector analysis}
\begin{table*}[!t]
 \centering
  \caption{STOI scores (\%) of different numbers of clean speech basis vectors $(\ddot J = 2,\overline{J} = 40,{\ddot{K}}=70)$}
  \label{tab:clean vector STOI}
  \centering
 \begin{tabular}{ccccc}
    \toprule
    \textbf{SNR(dB)} & \textbf{-5} & \textbf{0} & \textbf{5} & \textbf{10} \\
    \midrule
   \textbf{Noisy}& 51.59 ($\pm\,{0.50}$) &	64.26 ($\pm\, 0.52)$ & 76.30 ($\pm\, 0.47$) & 85.86 ($\pm\, 0.36)$ \\
    \midrule
    \textbf{T-NMF}& 55.85 ($\pm\,0.76$) &68.96 ($\pm\,0.73$) & 79.51 ($\pm\,0.59$) & 86.78 ($\pm\,0.40$) \\
    \midrule
    \textbf{NMF-HMM, ${\overline{K}}=5$}& {59.27} ($\pm\, 0.85$) & 69.98 ($\pm\, 0.69$) & 77.53 ($\pm\, 0.54$) & 82.81 ($\pm\ 0.44$) \\
    \midrule                   
    \textbf{NMF-HMM, ${\overline{K}}=10$}& \textbf{59.98} ($\pm\, 0.87$) & {71.94} ($\pm\,0.71 $) & {80.42} ($\pm\,0.53 $) & 86.20 ($\pm\, 0.38$)\\
    \midrule  
    \textbf{NMF-HMM, ${\overline{K}}=25$}& {59.84 ($\pm\,0.87$)} & \textbf{73.46 ($\pm\,0.75$)} & \textbf{83.17 ($\pm\,0.55$)} & \textbf{89.33 ($\pm\,0.36$)}\\
    \midrule  
    \textbf{NMF-HMM, ${\overline{K}}=50$}& {57.41 ($\pm\, 0.83$)} & {71.08 ($\pm\,0.77 $)} & {81.79 ($\pm\, 0.59$)} & {89.11 ($\pm\, 0.38$)}\\
    \bottomrule                             
  \end{tabular}
\end{table*}
\begin{table*}[!t]
 \centering
  \caption{STOI scores (\%) of different numbers of noise basis vectors $(\ddot J = 2,\overline{J} = 40,{\overline{K}}=25)$}
  \label{tab:noise vector STOI}
  \centering
  \begin{tabular}{ccccc}
    \toprule
    \textbf{SNR(dB)} & \textbf{-5} & \textbf{0} & \textbf{5} & \textbf{10} \\
    \midrule
   \textbf{Noisy}& 51.59 ($\pm\,{0.50}$) &	64.26 ($\pm\, 0.52)$ & 76.30 ($\pm\, 0.47$) & 85.86 ($\pm\, 0.36)$ \\
    \midrule
    \textbf{T-NMF}& 55.85 ($\pm\,0.76$) &68.96 ($\pm\,0.73$) & 79.51 ($\pm\,0.59$) & 86.78 ($\pm\,0.40$) \\
    \midrule
    \textbf{NMF-HMM, ${\ddot{K} = 10}$}& {55.74} ($\pm\, 0.79$) & 69.40 ($\pm\, 0.77$) & 80.80 ($\pm\, 0.60$) & {88.85} ($\pm\ 0.39$) \\
    \midrule                   
    \textbf{NMF-HMM, ${\ddot{K} = 20}$}& {57.28} ($\pm\, 0.82$) & {70.86} ($\pm\, 0.77$) & {81.69} ($\pm\, 0.58$) & {89.20} ($\pm\, 0.37$)\\
    \midrule  
    \textbf{NMF-HMM, ${\ddot{K} = 40}$}& 58.57 ($\pm\, 0.85$) & 71.99 ($\pm\, 0.76$) & {82.23} ($\pm\, 0.57$) & {89.22} ($\pm\, 0.36$)\\
    \midrule  
    \textbf{NMF-HMM, ${\ddot{K} = 70}$}& \textbf{59.84 ($\pm\,0.87$)} & \textbf{73.46 ($\pm\,0.75$)} & \textbf{83.17 ($\pm\,0.55$)} &\textbf{89.33($\pm\,0.36$)}\\
    
    \bottomrule                             
  \end{tabular}
\end{table*}

\begin{table*}[!btp]
 \centering
  \caption{Comparison of STOI scores (\%) for various algorithms under different SNRs using six types of noise.}
  \label{tab: Algorithm STOI}
  \centering
  
  \begin{tabular}{ccccccc}
    \toprule
    \textbf{SNR(dB)} & \textbf{Noise Type} & \textbf{Noisy} & \textbf{Log-MMSE} & \textbf{OMLSA} & \textbf{SLF-NMF} & \textbf{NMF-HMM} \\
    \midrule
    \centering
    \multirow{7}{*}{-5} & Babble & 48.28 & 44.64  & 43.89 & 47.81 & \textbf{49.95} 	\\
                        & F16 & 49.03 & 52.71 & 49.29 & 57.17 & \textbf{61.95}	\\
                        & Factory & 55.07 & 55.29 & 52.11 & 58.98 & \textbf{62.67}	\\
                        & destroyerengine & 51.37 & \textbf{57.01} & 51.62 & 50.83 & {51.81}	\\
                        & destroyerops & 54.09 & 55.00 & 52.23 & 54.18 & \textbf{56.99}	\\
                        & White & 53.97 & 55.76 & 48.95 & 58.38 & \textbf{64.79} \\	
	
    \cmidrule(r){2-7}                   
                       & Average &51.97 ($\pm\,0.08$) & 53.41 ($\pm\,0.09$) & 49.68 ($\pm\,0.09$) & 54.56 ($\pm\,0.09$) & \textbf{58.03} ($\pm\,0.11$) \\
    \midrule
    \multirow{7}{*}{0}  & Babble & 60.54 & 58.02 & 57.96 & 61.72 & \textbf{66.14} \\	
                        & F16 & 62.09 & 66.15 & 63.05 & 69.75 & \textbf{75.25}	 \\
                        & Factory & 67.64 & 67.85 & 66.15 & 71.69 & \textbf{75.48} \\	
                        & destroyerengine & 64.03 &  \textbf{69.95} & 65.22 & 63.36 &{67.26} \\	
                        & destroyerops & 65.49 & 66.71 & 64.34 & 66.24 & \textbf{70.75}\\
                        & White & 66.79 & 68.39 & 63.64 & 69.61 & \textbf{76.97}	 \\

    \cmidrule(r){2-7}                   
                        & Average & 66.43 ($\pm\,0.08$) & 66.17 ($\pm\,0.08$) & 63.39 ($\pm\,0.09$)  & 67.06 ($\pm\,0.09$) & \textbf{71.97} ($\pm\,0.09$) \\
    \midrule                   
    \multirow{7}{*}{5}  & Babble & 72.57 & 71.48 & 71.95 & 73.75 & \textbf{78.72}	\\
                        & F16 & 74.55 & 77.59 & 76.32 & 79.28 & \textbf{83.96}	\\
                        & Factory & 78.83 & 78.70 & 78.38 & 79.81 & \textbf{84.23}	\\
                        & destroyerengine & 76.67 & \textbf{80.86} & 78.29 & 75.95 & {80.43}	\\
                        & destroyerops & 75.72 & 76.84 & 74.98 & 76.98 & \textbf{80.89}\\
                        & White & 79.27 & 79.76 & 77.89 & 80.44 & \textbf{85.78} \\	

    \cmidrule(r){2-7}
                        & Average & 76.27 ($\pm\,0.07$) & 77.54 ($\pm\,0.07$) & 76.31 ($\pm\,0.08$) & 77.70 ($\pm\,0.07$) & \textbf{82.34} ($\pm\,0.07$)  \\

    \midrule               
    \multirow{7}{*}{10} & Babble & 82.66 & 81.98 & 83.08 & 83.05 & \textbf{86.96}	\\
                        & F16 & 84.88 & 86.40 & 86.53 & 86.49 & \textbf{89.43}	\\
                        & Factory & 87.02 & 86.40 & 87.16 & 87.31 & \textbf{89.38}\\
                        & destroyerengine & 86.92 & \textbf{88.80} & 88.17 & 85.99 & {88.30}	\\
                        & destroyerops & 84.18 & 84.76 & 83.86 & 84.63 & \textbf{87.58} \\	
                        & White & 88.88 & 88.14 & 87.95 & 89.05 & \textbf{91.51} \\	

    \cmidrule(r){2-7}
    
                         & Average & 85.76 ($\pm\,0.05$) & 86.08 ($\pm\,0.05$) & 86.13 ($\pm\,0.06$) & 86.09 ($\pm\,0.05$) & \textbf{88.87} ($\pm\,0.04$)\\
    \bottomrule                             
  \end{tabular}
\end{table*}



The number of basis vectors is another important parameter for proposed NMF-HMM algorithm. In this part, we investigate its influence to the PESQ and STOI performance. The number of clean speech states and noise states are set to $\overline{J} =40$ and $\ddot J = 2 $, respectively. First, we investigate the influence of the number of clean speech basis vectors. The number of noise basis vectors is set to $\ddot K = 70$. The number
of clean speech basis vectors ${\overline{K}}$ is set to 5, 10, 25, 50. The
other settings are set to the same as the previous experiment. Figure~\ref{fig:clean dimension} shows the average PESQ scores for different number of clean speech basis vectors. As can be seen, the proposed NMF-HMM method achieves a higher PESQ score than T-NMF in most cases. However, higher ${\overline{K}}$ cannot ensure better PESQ scores for the proposed algorithm probably due to the models overfitting to the limited training set. Additionally, the proposed method achieves the highest PESQ score when ${\overline{K}}=10$ at 0 and 5 dB SNRs. Table~\ref{tab:clean vector STOI} compares STOI scores for different ${\overline{K}}$. As can be seen, when ${\overline{K}}=25$,  the proposed NMF-HMM achieves the highest STOI score under 0 to 10 dB scenarios. In -5 dB SNR, the NMF-HMM with $\overline{K}=10$ has the best performance. However, the STOI scores of the NMF-HMM with $\overline{K}=25$ and $\overline{K}=10$ are comparable in -5 db scenario. From the PESQ and STOI results, it can be concluded that that a better PESQ can be obtained for the proposed NMF-HMM with ${\overline{K}}=10$ but a high STOI when ${\overline{K}}=25$. We set ${\overline{K}}=25$ for the subsequent experiments. 

In the next experiment, the effect of number of noise basis vectors is investigated. Here, we set $\overline{J}=40$, $\ddot{J}=2$, ${\overline{K}}=25$ and $\ddot K $ is set to 10, 20 ,40, and 70. The T-NMF is used as the reference method. Figure~\ref{fig:noise dimension} shows the average PESQ scores for different $\ddot K $. It can be seen that the a higher number of noise basis vectors leads to a higher PESQ score for both the T-NMF and NMF-HMM. In all SNR scenarios, the NMF-HMM with $\ddot{K}=70$ has the best PESQ performance. The average STOI scores with different $\ddot K $ are given in Table~\ref{tab:noise vector STOI}. As can be seen, the highest STOI score is obtained when ${\ddot{K}}=70$. Finally, based on the PESQ and STOI test results, we set ${\ddot{K}}=70$ for the subsequent experiments.

Combining above results and analysis, for the proposed NMF-HMM, we set ${\overline{J}=40,\ddot{J}=2,\overline{K}=25,\ddot{K}=70}$ to conduct the rest of experiments. It should be noted that this parameter settings may not be appropriate for a different training or test database. A good choice of parameters for the proposed algorithm is to balance the errors of overfitting and underfitting.

\subsection{Overall Evaluation}

In this section, we compare the proposed NMF-HMM speech enhancement method with state-of-the-art speech enhancement methods. We choose the optimally-modified log-spectral amplitude (OM-LSA) method \cite{cohen2001speech} with IMCRA noise estimator \cite{cohen2003noise}, span linear filters method \cite{jensen2015noise} (SLF-NMF), which applied the parametric NMF\cite{kavalekalam2018online} and Log-MMSE \cite{gerkmann2011unbiased} algorithm as the reference methods. For the SLF-NMF, the maximum SNR filter is applied. The codebook size of clean speech and noise is set to 64 and 8, respectively.   

\label{sub:overallcompare}
\begin{figure*}[!tbp]
  \centering
  \includegraphics[scale=0.8]{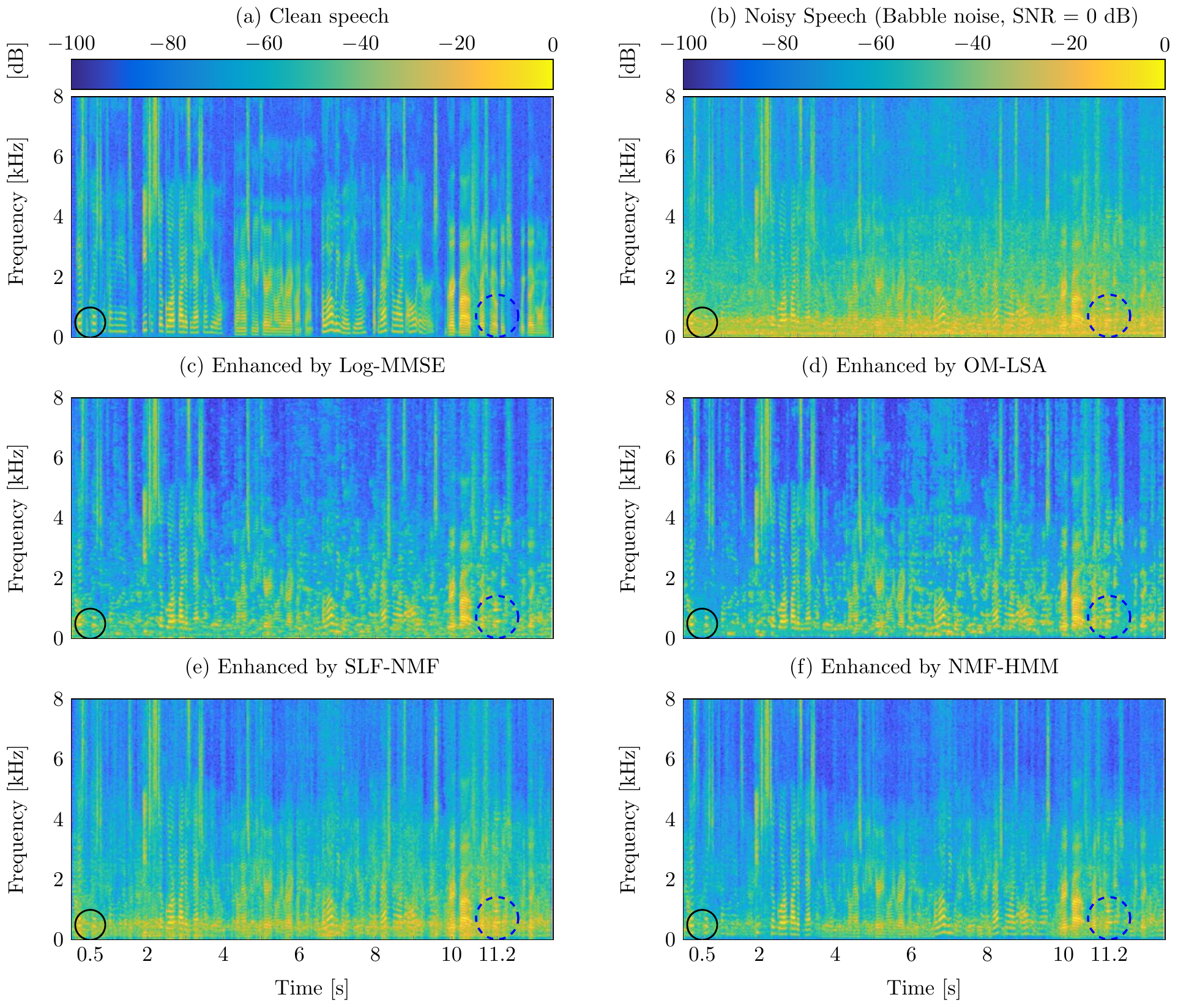}
  \centering
  \caption{Spectrograms of the clean speech, noisy speech and enhanced signals by various speech enhancement algorithms under 0 dB babble noise.}
  \centering
  \label{fig:spec}
\end{figure*}

First, the enhancement algorithms are used to process a noisy speech signal in babble noise under 0 dB SNR. The spectrograms of the clean speech, noisy signal and enhanced signals are shown in Figure~\ref{fig:spec}. It can be seen that all the methods enhances the speech efficiently. However, the unsupervised methods (OMLSA and Log-MMSE) remove more background noise, but they also introduce musical noise, degrading the intelligibility. The supervised NMF-based methods (SLF-NMF and NMF-HMM) can recover more speech characteristics than the unsupervised methods, especially in the low-frequency area. Comparing NMF-HMM with other methods, it can be found that NMF-HMM can recover a better harmonic structure. For example, a better recovery of the harmonic structure of the proposed NMF-HMM can be seen at 0.5 and 11.2 seconds (inside the black and blue circles). Moreover, compared with the SLF-NMF, the NMF-HMM removes more noise between the harmonic bands (e.g., the area inside the black circle). Furthermore, the proposed NMF-HMM does not generate musical noise. Therefore, comparing with the reference methods, the best speech enhancement performance can be obtained using the proposed NMF-HMM.

\begin{figure}[!tbp]
  \centering
  \includegraphics[width=\linewidth]{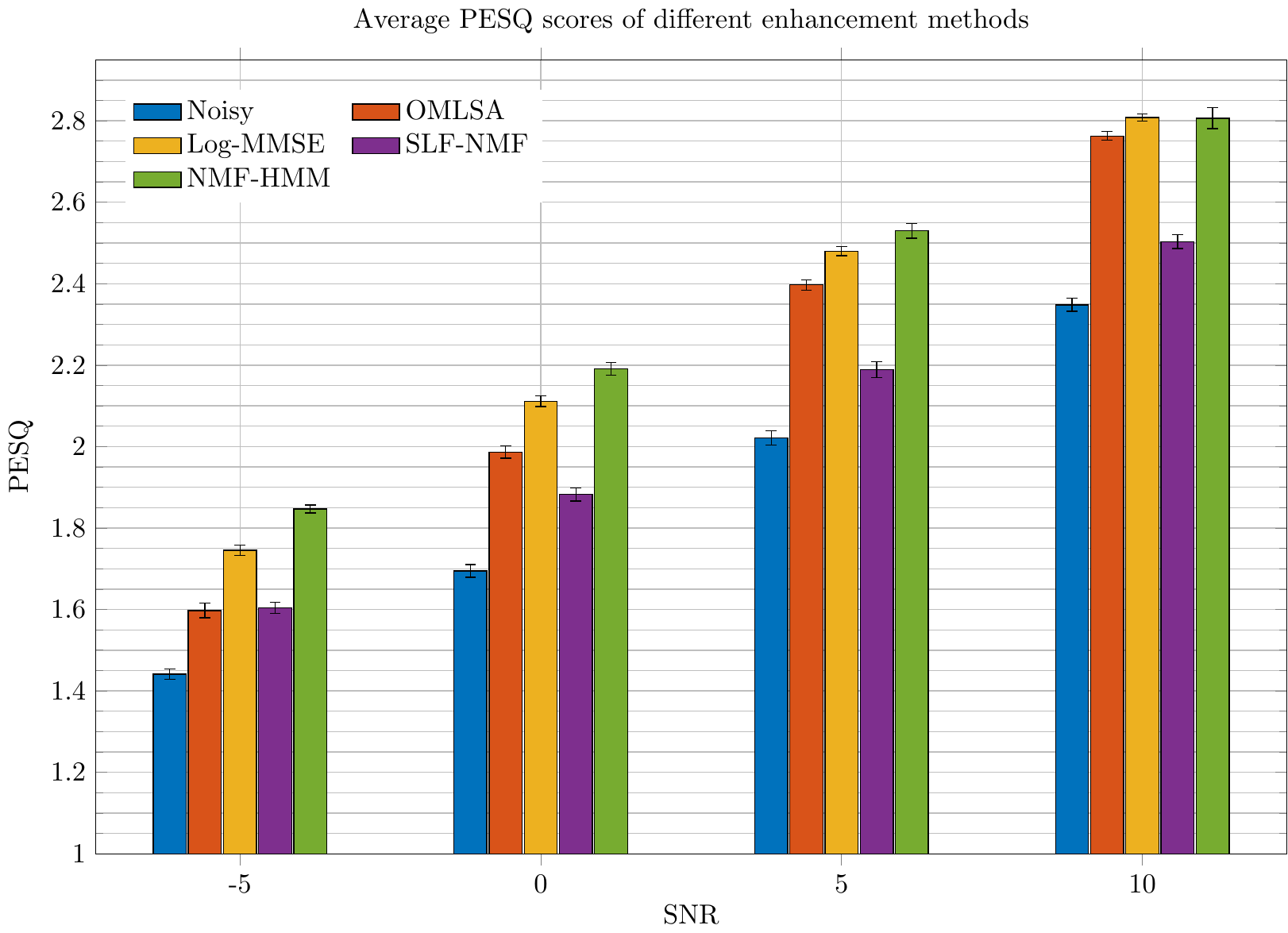}
  \caption{Average PESQ scores of different algorithms using six types of noise under different SNRs.}
  \label{fig:average PESQ }
\end{figure}

Second, the performance of the proposed NMF-HMM, SLF-NMF, Log-MMSE, and OM-LSA are evaluated using the test set.  It should be noted that the destroyerengine and destroyerops noise, not in the training set, are included in the test set to evaluate the performance.  Figure~\ref{fig:average PESQ } shows the average PESQ scores with 95\% confidence interval of these algorithms. As can be seen, the SLF-NMF has the worst performance among these algorithms. This is due to its poor performance for dealing with unseen noise types. Moreover, the proposed NMF-HMM outperforms other enhancement algorithms in all SNR scenarios, except that the Log-MMSE and NMF-HMM have a similar performance when SNR=10. Furthermore, in low SNR scenarios (e.g., -5--5 dB), the average PESQ score improvement of the proposed NMF-HMM is larger than 0.05 against the other algorithms. The results of the STOI scores with 95\% confidence interval for various algorithms are given in Table V. As can be seen, the SLF-NMF has a better performance than the Log-MMSE and OMLSA when the noise types are in the training set. However, under unseen noise cases, that is destroyerengine and destroyerops noise, the performance of the SLF-NMF is comparable to the OMLSA.  The Log-MMSE has the best performance in all cases for the destroyerengine noise. The proposed NMF-HMM achieves the best speech enhancement performance in all the other cases. Moreover, in high SNRs (e.g, 5-10 dB) under destroyerengine noise, the STOI scores of the  proposed NMF-HMM  and Log-MMSE are comparable. Furthermore, compared with the reference methods, the NMF-HMM has the highest average STOI score. 

\section{Conclusion}

In this work, we propose a NMF-HMM-based speech enhancement algorithm that applies the sum of Poisson, leading to the KL divergence measure, as the observation model for each state of HMM. The computationally efficient multiplicative update rule can be used to conduct parameter updates during the training stage for the proposed method. Moreover, using the HMM, the temporal dynamic information of speech signals can  be captured. Furthermore, we also propose a novel NMF-HMM-based MMSE estimator to conduct online speech enhancement. The parallel computation can be applied for the proposed estimator, so we can effectively reduce the time complexity during the online speech enhancement stage.  By experiments, suitable number of  states basis vectors for the proposed NMF-HMM are found.  Our experimental results also indicate that the proposed algorithm can outperform state-of-the-art NMF-based and unsupervised speech enhancement methods. In addition, the NMF-HMM can achieve the better performance than SLF-NMF in the tested unseen noise cases (destroyerengine and destroyerops). The generalization ability of the proposed method in other unseen noise cases is expected to be investigated in the future work.


%

\appendices




\ifCLASSOPTIONcaptionsoff
  \newpage
\fi



%

\bibliographystyle{IEEEtran}
\bibliography{IEEEabrv,myabrv,IEEEexample}

%








\end{document}